\documentclass[aps,pra,amsfonts,longbibliography]{revtex4-2}
\usepackage{bm,amsmath,amssymb,graphicx,tikz,epstopdf}
\usepackage{xcolor,hyperref,subeqnarray}
\def \bm#1{{\bf #1}}
\def\r{{\bf r}}

\begin{document}
	\title{Microswimmers in  vortices: Dynamics and trapping}

\author{Ivan Tanasijevi\'c}

\email[]{it279@cam.ac.uk}

\affiliation{Department of Applied Mathematics and Theoretical Physics, Centre for Mathematical Sciences, University of Cambridge, Wilberforce Road, Cambridge CB3 0WA, United Kingdom}

\author{Eric Lauga}

\email[]{e.lauga@damtp.cam.ac.uk}

\affiliation{Department of Applied Mathematics and Theoretical Physics, Centre for Mathematical Sciences, University of Cambridge, Wilberforce Road, Cambridge CB3 0WA, United Kingdom}

\begin{abstract}
	
	Biological and artificial microswimmers often self-propel  in  external  flows of vortical nature; relevant  examples include algae in small-scale ocean eddies, spermatozoa in uterine peristaltic flows and bacteria in microfluidic devices. A recent experiment has shown that  swimming bacteria in model vortices are expelled from the vortex all the way to a well-defined depletion zone (Sokolov and Aranson (2016) ``Rapid expulsion of microswimmers by a vortical flow.''{\it Nature Comm.} {\bf 7}, 11114). In this paper, we propose a  theoretical model to investigate the dynamics of elongated microswimmers in elementary vortices, namely active particles in  two- and three-dimensional  rotlets.  A deterministic model first reveals the existence of bounded orbits near the centre of the vortex and unbounded orbits elsewhere. We further discover a conserved quantity of motion that allows us to map the phase space according to the type of the orbit (bounded vs unbounded). We next introduce  translational and rotational  noise into the system. Using a Fokker--Planck formalism, we quantify the quality of trapping near the centre of the vortex by examining the probability of escape and the mean time of escape from the region of deterministically bounded orbits. We finally show how to use these findings to formulate a prediction for the radius of the depletion zone, which compares favourably with the experiments  of Sokolov and Aranson (2016).
	
\end{abstract}

\date{\today}

\maketitle

\section{Introduction}

Motile microorganisms, ubiquitous in nature, have been shown to exploit a wide range of physical mechanisms to self propel  through their fluid environment~\cite{lighthill75,lauga_book}. These swimming microorganisms must not only ensure successful propulsion but also navigate against   external flows~\cite{StokerReviewBacteriaPlankton, fauci06,StokerReviewPlanktons, PedleyReview,rusconi2015microbes}. The novel dynamics that arises from the interaction of motile microorganisms and external flows is commonly known as the {\it rheotaxis}.

Perhaps the most famous example of rheotaxis is the hydrodynamic focusing of bottom-heavy algae in downward flows as observed  in both laboratory experiments and oceanic flows~\cite{Kessler1985, pedley_kessler_1990,GyroTurbulence1,GyroTurbulence2,GyroTurbulence3}. This focusing results from a combination of  hydrodynamic interactions with   external flows and the gravitational alignment of  the bottom-heavy algae with the vertical direction; as a result, it is often known as  {\it gyrotaxis}. 
Further examples of motile cells that show a  rheotactic response include bacteria~\cite{EcoliJeffrey,EcoliRheotaxis,BacterialRheotaxis,EcoliAnkeLindner,BacterialCrystalDispersion} and mammalian spermatozoa~\cite{sperm1,SpermRayRheot,SpermShear,SpermReview,miki2013rheotaxis}. 
Unsurprisingly,  bio-inspired artificial  microswimmers are faced with similar rheotactic challenges~\cite{MicroNanoBook,ArtificialRheotaxis,ArtificialRodRheotaxis,RheotaxisBimetalicMicro,Potomkin_2017,RheotaxisArtificialSphere}.

To understand the essential physics of  rheotaxis, minimal mathematical models of motile cells in flows have recently been proposed~\cite{RamaswamyActiveParticles,TorneyNeufeld2007,ten_Hagen_2011,Stark2016}. 
In the simplest model, the swimmer is an active particle of fixed shape advected by the flow while swimming with a prescribed  speed along a  fixed direction in the swimmer's frame. The flow changes not only the location of the swimmer (advection) but also its swimming direction (reorientation).  For spherical swimmers, the minimal model can have closed-form solutions, as long as the swimmer is assumed to move through an infinite fluid. Indeed, in that case the equations of motion are exact, with interactions with the flow  following from  Fax\'en's laws~\cite{KimKarrila}. Fundamental  solutions exist then for swimming in simple linear flows such as the solid-body rotation, shear and extensional flows~\cite{lauga_book,linearflow}. 
In the  more complex case of swimming in a Poiseuille flow, but  perhaps one more relevant to applications, the dynamics of a spherical swimmer cannot be exactly integrated but it was shown to have an underlying Hamiltonian structure~\cite{poisseuille}.

A more realistic version of the minimal model consider the swimmers to have the shapes of elongated spheroids. This allows the model to capture phenomena arising from the elongated  form of real cells, as relevant for example for flagellated bacteria and spermatozoa~\cite{lauga_book}. In turn, this assumption makes analytical predictions more complex unless  additional assumptions are made.  
Focusing on flows that have  a typical gradient length scale much larger than the size of the swimmer, the equations of motion can be approximated using the classical  {Jeffery's} equations~\cite{Jeffery} that exactly describe the behaviour of passive spheroids in a shear flow. 	 
{For example,   eddies in the ocean relevant for microorganisms have sizes on the order of  millimetres, thus always much larger than the    
	micro-algae that populate them~(tens of microns).} 	 The validity of Jeffery's equations as applied to elongated biological organisms has been further verified using direct experiments with the bacterium {\it E.~coli} in microfluidic channels~\cite{EcoliJeffrey}. {Note that using  {Jeffery's} equations, classical work has been carried out on the dynamics of passive elongated particles in external flows~\cite{cylindrical_particles,noise_elongated,leal_hinch_1971_shear} as well as some recent work on active elongated swimmers in linear~\cite{HyperbolicFlow} and pressure-driven flows~\cite{poisseuille,PressureDriven}.}	
In a Poiseuille flow, and in contrast to the case of a spherical swimmer, the equations of motion for an elongated swimmer no longer have the Hamiltonian structure  but a conserved quantity still exists~\cite{zottl2013periodic} and two types of trajectories are seen in this case: upstream swinging motion around the middle of the channel or   downstream tumbling closer to the channel walls. Both types of trajectories were confirmed experimentally  for motile bacteria~\cite{rusconi14}.

In this work, we focus on the behaviour of elongated microswimmers in elementary vortical flows. A well-known example where biological microswimmers  have to self-propel in vortices are algae and bacteria swimming in small-scale ocean eddies~\cite{khurana2011reduced,khurana2012interactions,TurbulenceVortex1,TurbulenceVortex2}. Other relevant examples include spermatozoa swimming in	uterine peristaltic flows~\cite{fauci06} and  recirculation flows generically occurring in  standard microfluidic devices~\cite{cavity,Li2021}.    Motivated by this fundamental problem, a recent experimental study on swimming bacteria   reported the existence of a cell depletion zone  in a vortical flow created by an externally rotated body~\cite{depletion}.   This depletion zone  forms around the body after it is forced into rotation in an otherwise uniform, dilute suspension of swimming bacteria. In contrast,  recent theoretical work predicted a zone around the centre of a vortex with bounded orbits~\cite{Arguedas_Leiva_2020}.

In this paper,  we use a theoretical treatment to   reconcile the experimentally-observed depletion~\cite{depletion} with the theoretically-predicted trapping~\cite{Arguedas_Leiva_2020}. 	 Specifically, we consider  the dynamics of model swimmers in two- and three-dimensional   rotlets, i.e.~the  rotational   flows that exactly represent the flow of rotating bodies (cylinders and spheres) in a bulk fluid at low Reynolds number~\cite{KimKarrila} (setup described in \S~\ref{sec::deterministic}).
We  mathematically determine the trapped orbits of swimmers in both types of singular vortices, in the absence of any noise in the system (spherical swimmers in \S~\ref{sec::spherical} and elongated swimmers in  \S~\ref{sec:trappingprolate}). Next, we include translation and rotation noise and show how to quantify  trapping in these vortices (\S~\ref{noise}). Finally, we   use our mathematical model to formulate a prediction for the radius of the depletion zone, which we show compares favourably with  experimental observations in Ref.~\cite{depletion} (\S~\ref{sec:depletion}).

\section{Setup and  deterministic model \label{sec::deterministic}}

In order to address the physical behaviour of swimming cells in external flows, we consider the fundamental elongated `active particle' model.  	A swimming cell is modelled as a prolate spheroid, of fixed aspect ratio $a\geq1$ and major axis $d$, which is being advected and rotated by the steady external flow $\bm{u}(\bm{r})$. In addition to the flow advection, the cell   swims with velocity $V_0 \bm{p}$ of fixed magnitude $V_0$ in the direction of the major axis of the spheroid (unit vector $\bm{p}$ in that direction). 	Using the aforementioned Fax\'en's laws and neglecting the  size $d$ of the swimmer relative to the typical length scale characterising the flow gradients~\cite{lauga_book}, the  position $\r (t)$ of the cell evolves in time as
\begin{equation} \label{eq::position}
	\dot{\bm{r}} = \bm{u}(\bm{r})+ V_0 \bm{p}.
\end{equation}

Following the classical result of Jeffery~\cite{Jeffery}, and using the same assumption on the size of the swimmer,  we may model the rotation of the swimmer as that of a passive spheroid in a linear flow, given by the equation
\begin{equation}
	\frac{d \bm{p}}{dt} = \frac{1}{2} \pmb{\Omega}\times\bm{p}+B(\mathbf{I}-\bm{p}\bm{p})\cdot \bm{E}\cdot \bm{p}, \label{eq::jeffrey}
\end{equation}  
where $\mathbf{I}$ is the identity tensor, $\bm{E} = (\bm{\nabla}\bm{u}+\bm{\nabla}\bm{u}^{T})/2$ is the symmetric rate of strain tensor  of the external flow, $\pmb{\Omega} = \bm{\nabla}\times \bm{u}$ is its vorticity and $B = (a^2-1)/(a^2+1) \in [0,1)$ is the swimmer's shape factor ($B=0$  for a sphere and $B\rightarrow 1$ for an elongated rod). 
Using this fundamental active particle model, we  investigate below the dynamics of  swimmers in  external vortical flows.

\section{Spherical swimmers in vortices \label{sec::spherical}}
We start by the simplest case of spherical swimmers, i.e.~with $B=0$ in Eq.~\eqref{eq::jeffrey}. We  mathematically demonstrate the existence of two classes of deterministic trajectories, bounded and unbounded, using a theoretical approach that will be exploited further in the case of elongated cells ($B>0$, \S~\ref{sec:trappingprolate}). We focus on the case where the flow is  the   three-dimensional (3D) Stokes flow  $\bm{u}$ created by a sphere of radius $R$ rotating with angular velocity $\pmb{\omega} = \omega \hat{\bm{z}}$, namely
\begin{equation}
	\bm{u} =  \frac{R^3}{r^3} \pmb{\omega}\times \bm{r},
\end{equation}
where $r = |\bm{r}|$. This solution is also known at the  3D rotlet, i.e.~the flow created by a point torque ${\bf L} = 8\pi \mu R^3 {\pmb \omega}$ located at the origin. { In later sections of this paper, we will investigate the two-dimensional (2D) version of this singular vortex flow, i.e.~the 2D rotlet. That flow can be realised as the Stokes flow around a rotating, infinitely long cylinder. However, since the 2D rotlet    has no vorticity (see \S \ref{sec:prolate2D}), it does not affect the dynamics of spherical swimmers, hence our focus on the 3D case here.}

\begin{figure}
	\includegraphics[width=0.85\linewidth]{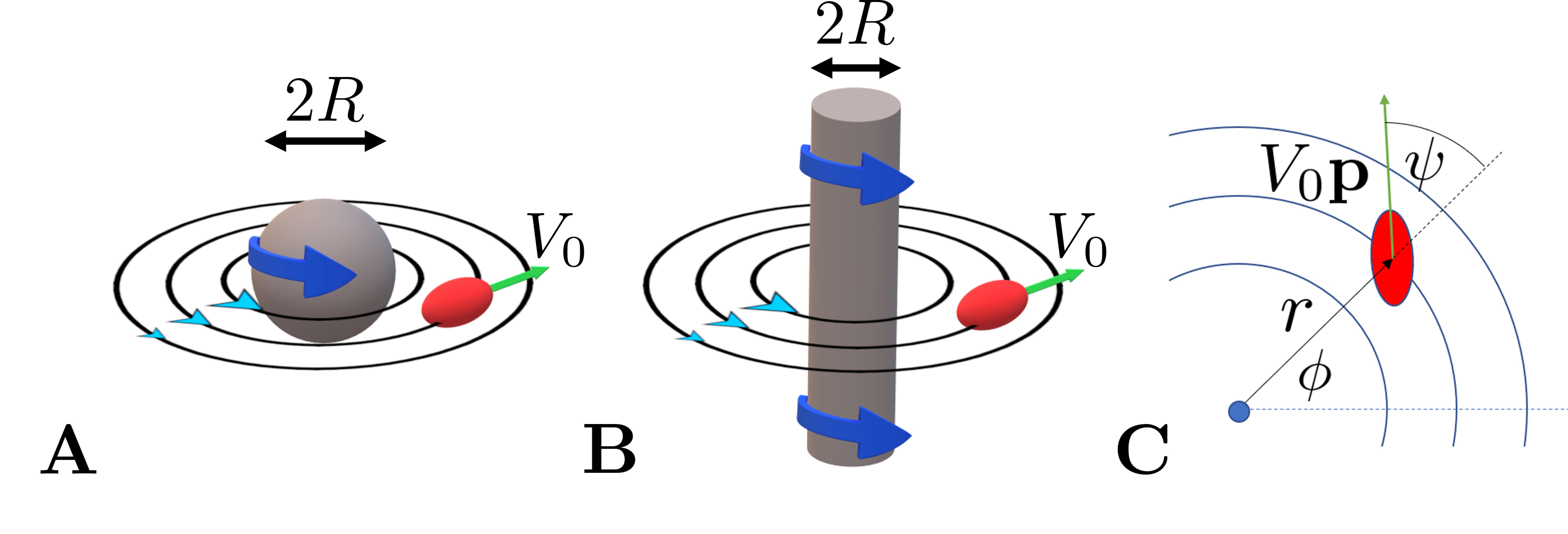} 
	\caption{Illustration of the model swimmer in a 3D rotlet flow, i.e.~the flow outside a rotating sphere (A), and  in a 2D rotlet flow outside  a rotating cylinder (B). The swimmer is sketched as a red spheroid and black lines show the flow streamlines. C: Notation used throughout this paper with the location of the swimmer described using 		planar polar coordinates $(r,\phi)$  and its orientation using the angle  $\psi$.} \label{fig::setup}
\end{figure}
We confine the swimmer to move in the sphere's equatorial plane (through its centre and perpendicular to $\hat{\bm{z}}$, see illustration in Fig.~\ref{fig::setup}A).  
We describe the location of the swimmer in the plane using the planar polar coordinates $(r,\phi)$  and its orientation using the angle  $\psi$ between its orientation vector $\bm{p}$ and  the radial unit vector  $\bm{r}/r$ at its position (see Fig.~\ref{fig::setup}C).

Using this notation, the equations of motion, Eqs.~\eqref{eq::position}-\eqref{eq::jeffrey}, 
take the   form
\begin{subeqnarray}
	\dot{r} &=& V_0 \cos \psi, \slabel{eq3dsphere1}\\
	\dot{\phi} &=& \frac{V_0}{r} \sin \psi +\frac{\omega R^3}{r^3}, \slabel{eq3dsphere2}\\ 
	\dot{\psi} &=& -\frac{3\omega R^3}{2r^3} -\frac{V_0}{r}\sin \psi, \slabel{eq3dsphere3}
\end{subeqnarray}
After nondimensionalising these equations using $R$ as the relevant length scale and $\omega^{-1}$ as  time scale, a straightforward  manipulation of  Eqs.~\eqref{eq3dsphere1} and \eqref{eq3dsphere3} show that there exists  a  constant of motion in this dynamical system. Specifically, if we define
\begin{equation}\label{conserv}
	h \triangleq \alpha r \sin \psi -\frac{3}{2r},
\end{equation}
where $\alpha = V_0/\omega R$ is the non-dimensional swimming speed, we see that $\dot h=0$.  

{Next,
	using the fact that $\dot{r}^2 = \alpha^2(1-\sin^2 \psi)$ we can express $\sin \psi(r) = h/\alpha r+3/2\alpha r^2$ from Eq.~\eqref{conserv} to notice that 
	\begin{equation}
		\frac{\dot{r}^2}{2}+\frac{\alpha^2}{2}\left[\left(\frac{h}{\alpha}+\frac{3}{2\alpha r^2}\right)^2-1\right]=0,
	\end{equation}
	and thus $r$ behaves }
as if it was under the influence of an effective potential 
\begin{equation}
	V(r) = \frac{1}{2}\left[\left(\frac{h}{r}+\frac{3}{2 r^2}\right)^2-\alpha^2 \right],
\end{equation} 
with an energy-like quantity $E = \dot{r}^2 /2 +V(r)$ equal to 0 for all times. 

Since the effective potential $V$ has limits $V\rightarrow -\alpha^2 /2$ as $r\rightarrow \infty$ and   $V\rightarrow +\infty$ as $r\rightarrow 0$,   the entrapment relies on the existence of a local maximum of $V(r)>0$ to prevent the swimmer from escaping to infinity. Taking the derivative, we see that the condition $dV/dr =0 $ 
is equivalent to $r h = -3/2$ or $r h = -3$. Clearly, if $h\geq 0$, then the potential is monotonic and the swimmer escapes. On the other hand, if $h<0$, then we obtain that at $r=-3/2h$ we have a minimum of the potential, with $V_{\rm min}=-\alpha^2 /2$ while at $r=-3/h$ there is a local maximum  $V_{\rm max} = (h^4-36\alpha^2)/72$. Thus, we predict theoretically that the swimmer will be trapped if and only if $r_0<-3/h$ and $V_{\rm max} > 0$ so that $h<-\sqrt{6\alpha}$. 
\begin{figure}[t]
	\includegraphics[width=0.7\linewidth]{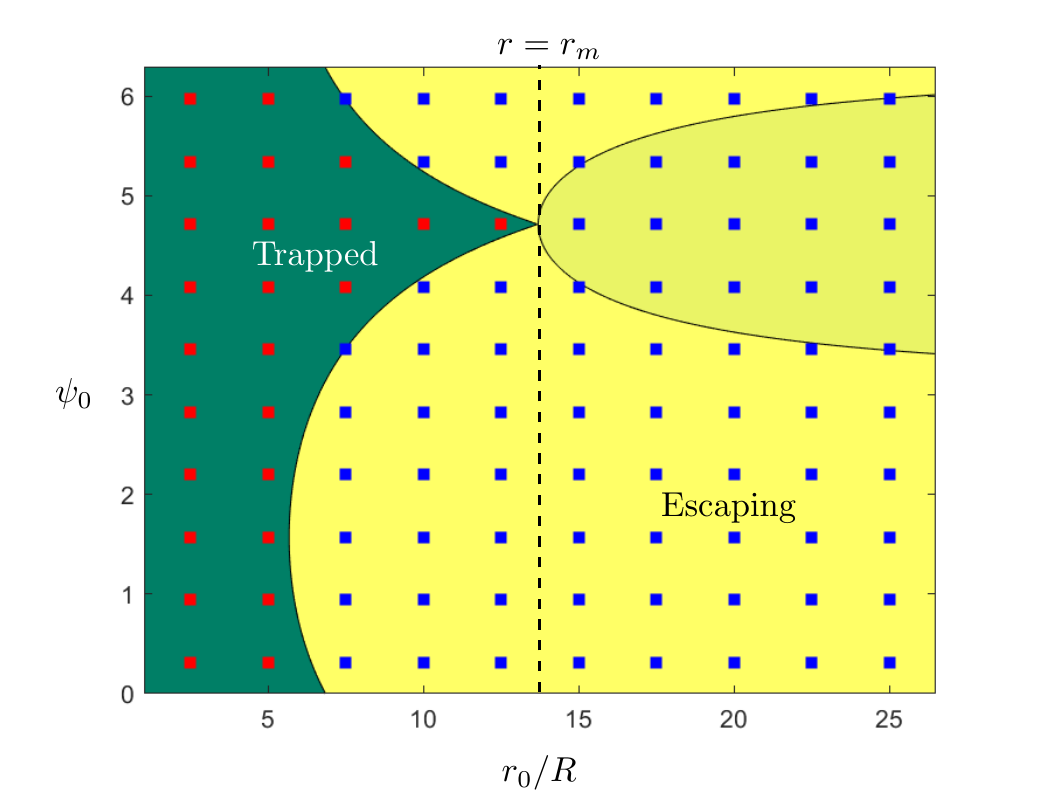}
	\caption{Trapping of a spherical swimmer in a 3D rotlet: a map of the parameter space of the initial position ($r_0/R$) and orientation angle ($\psi_0$), in the case $\alpha=0.008$. Yellow and green regions indicate theoretical predictions of open and bounded orbits, respectively. Shaded and unshaded yellow regions are only there to illustrate different theoretical conditions for escape{, namely $h<-\sqrt{6\alpha}$ (shaded) and $r_0<-3/h$ (unshaded).} 
		Square symbols represent the outcomes of numerical simulations: red -- bounded orbits; blue -- open orbits.}
	\label{trapfig}
\end{figure}

{ 
	These   theoretical predictions can be used to validate direct finite-difference simulations of the equations of motion in Eqs.~\eqref{eq3dsphere1}-\eqref{eq3dsphere3}; these simulations will then be used in the rest of the paper when exact theoretical predictions are harder to make}. Results   are shown in Fig.~\ref{trapfig} and we obtain excellent agreement between simulations and theoretical predictions. 
Using the initial position and orientation of the swimmer $(r_0, \psi_0)$, we note that the theoretical  conditions condense to a single equation, $r_0(\sqrt{1+\sin \psi_0}+1)<(3/2\alpha)^{1/2}$; this is the equation of the  separatrix of this dynamical system, i.e.~the interface between the light yellow and dark green regions in Fig.~\ref{trapfig}. An important feature of this phase map that we will use later  is the maximal radial distance $r = r_m$ for which a trapped state exists, given by $r_m = (3/2\alpha)^{1/2}$.

In  a recent theoretical study~\cite{Arguedas_Leiva_2020}, { based on the same model equations as the current work,} it has been noted that in a general axisymmetric flow of the form $u(r)\bm{e}_{\phi}$ the equations of motion for a spherical swimmer can be transformed to take a Hamiltonian
form. Indeed, under the transformation
\begin{subeqnarray}
	X &=& x \cos \phi + y \sin \phi, \\
	P &=& - x \sin \phi + y \cos \phi,
\end{subeqnarray} 
then Eqs.~\eqref{eq3dsphere1} and~\eqref{eq3dsphere2} become equivalent to a set of Hamilton's equations
{
	\begin{subeqnarray}
		\dot{X} &=& \frac{\partial H}{\partial P}, \\
		\dot{P} &=& -\frac{\partial H}{\partial X},
	\end{subeqnarray}
}
for the ``position'' $X$ and the ``momentum'' $P$ with the  Hamiltonian $H$ given by
\begin{equation}
	H = \alpha P +\int_{\infty}^{r} \frac{s^2}{2}\frac{d}{ds}\left[\frac{u(s)}{s}\right]\, ds.
\end{equation}
Since {the Hamiltonian $H$}  
has no explicit time dependence, it is a conserved quantity of motion and so we expect it to be a function of the conserved quantity $h$   derived above. For the case of a 3D rotlet that we consider here, it is simple to integrate and express $H$ as a function of $r$ and $\psi$ only as
\begin{equation}
	H = \alpha r \sin \psi -\int_{\infty}^{r}\frac{s^2}{2}\frac{d}{ds}\left[\frac{1}{s^3}\right]\, ds =  \alpha r \sin \psi - \frac{3}{2r}.
\end{equation}
In other words, $H=h$, so the Hamiltonian is equal to the conserved quantity    found above in Eq.~\eqref{conserv}.

\section{Elongated swimmers in vortices}
\label{sec:trappingprolate}

So far we only considered the case of spherical swimmers. In contrast, most biological or artificial swimmers have 
anisotropic shapes and are often elongated in the direction of swimming.	 	 When in an external flow, spherical swimmers  are not influenced by the local rate of strain of the fluid flow, while an elongated swimmer   experiences an additional torque aligning  it with the principal axes of  the local rate of strain. 	 Thus, the rotational dynamics can be drastically different for elongated swimmers, which could  lead to  qualitatively different behaviour in external flows. Therefore, we now  consider  the case of elongated swimmers in vortices. By deriving  the conserved quantity of motion for elongated swimmers in 3D and 2D rotlet flows, we show the existence of a  similar phase diagram as for spherical swimmers, but with a different separatrix between the region of trapped and unbounded orbits.

\subsection{Prolate spheroid in the equatorial plane of a 3D rotlet}
\begin{figure*}[t]
	\includegraphics[width=\linewidth]{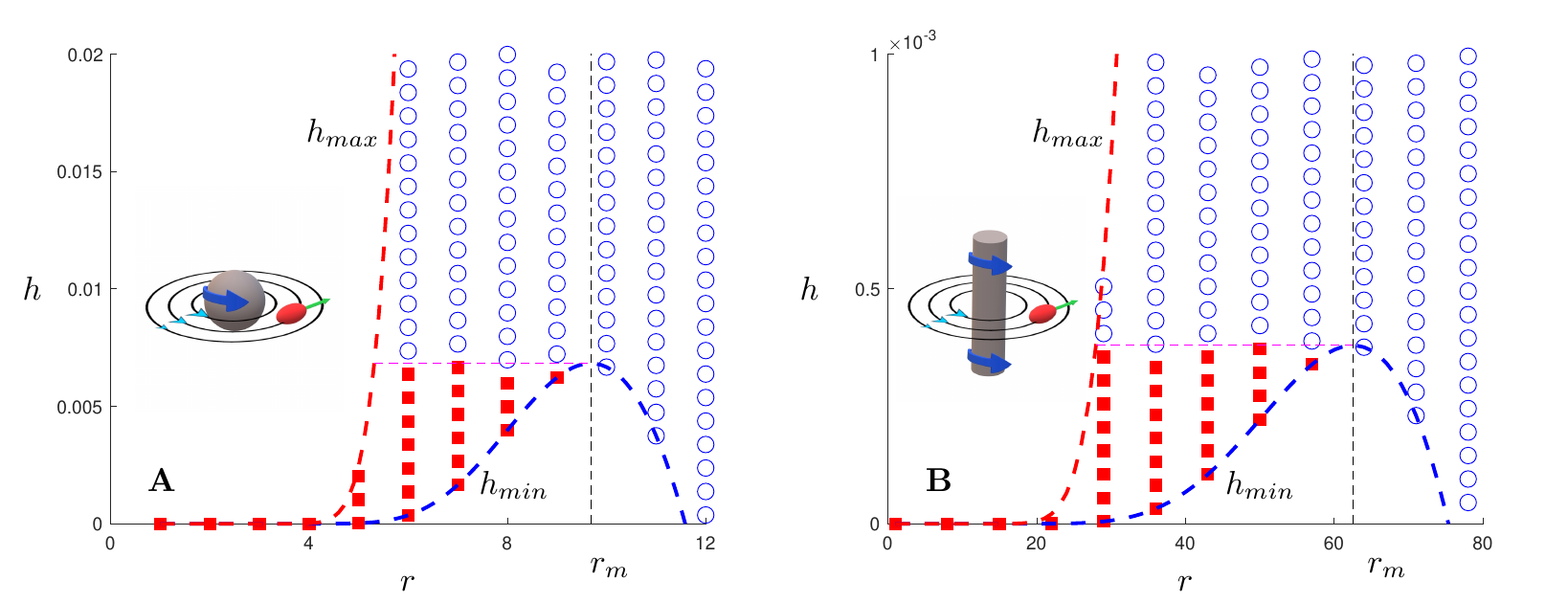}
	\caption{Trapping of elongated swimmers in a 3D rotlet flow (A) and a 2D rotlet (B). In both cases, we display a map in the $r-h$ parameter space in the case $B=0.5$ (corresponding to a swimmer of aspect ratio $a=\sqrt{3}\approx 1.73$)			and $\alpha=0.008$. Scattered symbols represent the outcomes of numerical simulations; red filled squares stand for bounded orbits and blue empty circles for unbounded ones. The dashed red and blue lines represent, respectively, the functions $h_{max}(r)$ and $h_{min}(r)$ derived analytically; the dashed black line shows the location of $r = r_m$ while the dashed pink line shows $h = h_{min}(r_m)$.}
	\label{rhspace}
\end{figure*}

Let the swimmer be a prolate spheroid of aspect ratio $a\geq 1$ swimming in the same external flow as above, i.e.~$\bm{u} = \pmb{\omega}\times \bm{r} R^3/r^3$. Using same notation as above, the non-dimensional equations of motion become 
\begin{subeqnarray}
	\dot{r} &=& \alpha \cos \psi, \slabel{eq::3dprolate1}\\
	\dot{\phi} &=& \frac{\alpha}{r} \sin \psi +\frac{1}{r^3}, \slabel{eq::3dprolate2} \\
	\dot{\psi} &=& -\frac{3}{2r^3}(1+B \cos 2\psi) -\frac{\alpha}{r}\sin \psi,\slabel{eq::3dprolate3}
\end{subeqnarray}
where we recall that $B=(a^2-1)/(a^2+1)\in (0,1)$ is the shape factor of the spheroid. If we introduce $u = \alpha r \sin \psi$ we can combine the radial and $\psi$ equations of motion to obtain
\begin{equation}
	\dot{u} = -\frac{3}{2 r^2}\dot{r}\left(1+B-2B\frac{u^2}{\alpha^2 r^2}\right).
\end{equation}
By introducing $x = B\alpha^{-2} r^{-3}$ and solving for $u(x)$, the equation above becomes 
\begin{equation}
	\frac{du}{dx} = -u^2 +\beta x^{-2/3},
\end{equation}
where $\beta = (1+B)\alpha^{2/3}B^{-1/3}/2$. This is a classical Riccati equation and we can transform it to the modified Bessel equation by first introducing $s(x)$ such that $u(x) = \frac{ds}{dx}/s$, then defining $z = 3\beta^{1/2} x^{3/2}/2$ and $\omega(z)$ such that $s(x) = z^{3/4} \omega(z)$. This transformation leaves us with solving the modified Bessel equation 
\begin{equation}
	z^2 \omega^{\prime\prime} +z\omega^\prime-(z^2+9/16)\omega = 0,
\end{equation}
whose solutions are $\omega(z) = c_1 I_{3/4}(z)+c_2 K_{3/4}(z)$, where $I, K$ are the modified Bessel functions of the first kind and $c_1$ and $c_2$ are constants. Hence we obtain
\begin{equation}
	\sin \psi (r) = \frac{\beta^{1/2}}{(\alpha B)^{1/3}}\frac{c_1 I_{-1/4}(z)-c_2 K_{-1/4}(z)}{c_1 I_{3/4}(z)+c_2 K_{3/4}(z)},
\end{equation}
where $z = 3\beta^{1/2} B^{2/3} \alpha^{-4/3} r^{-2}/2$. Then, $h=c_1/c_2$ is a conserved quantity of the motion and it is expressed as
\begin{equation}\label{hdef}
	h = \frac{\gamma K_{-1/4}(z)+\sin \psi K_{3/4}(z)}{\gamma I_{-1/4}(z)-\sin \psi I_{3/4}(z)},
\end{equation}
where $\gamma =  \beta^{1/2}(\alpha B)^{-1/3} = ((B+1)/2B)^{1/2}\geq1 $. {Note that $\gamma = a/\sqrt{a^2-1}$ represents the inverse of the eccentricity of the elliptical cross-section of the swimmer.
	
	Finally, we may now investigate the conditions for entrapment in case of a prolate swimmer in the equatorial plane of a 3D rotlet. For comparison, we perform agent based simulations of swimmers starting from various initial conditions $(r,\psi)$ (or equivalently $(r,h)$) and plot the symbols in Fig.~\ref{rhspace}A according to the type of the orbit that we find:  blue empty circles represent unbounded orbits while and red filled squares are used for bounded ones.} As opposed to the spherical case, it is less straightforward to examine the features of the effective radial potential analytically, so we introduce a different approach by considering   orbits in the $r-h$ space. Swimmers will clearly be moving along the constant-$h$ lines but not all of the $r-h$ space is physically feasible. Inspecting   Eq.~\eqref{hdef} we can see that for a given $r$ (i.e.~a given $z$), $h$ can only vary between $h_{min}(r)$ and $h_{max}(r)$, where
\begin{subeqnarray}
	h_{min}(r)&=&\frac{\gamma K_{-1/4}(z)- K_{3/4}(z)}{\gamma I_{-1/4}(z)+ I_{3/4}(z)}, \\
	h_{max}(r)&=&\frac{\gamma K_{-1/4}(z) + K_{3/4}(z)}{\gamma I_{-1/4}(z)- I_{3/4}(z)}.
\end{subeqnarray}
We claim that a swimmer is certain to be trapped if it starts from a point in the $r-h$ space such that its constant-$h$ line has two intersections with any of the  $h = h_{min}(r)$ or $h = h_{max}(r)$ lines (see illustration in Fig.~\ref{rhspace}A). 
Otherwise, it will escape due to the fact that $\dot{r} \propto \cos \psi$. Specifically, if the swimmer starts with $\cos \psi>0$ it will always advance in $r$ since $\dot{r} \propto \cos \psi$ and the change in the sign of $\cos \psi$ can only happen when $\sin \psi = \pm 1$, which are   exactly the $h_{min}$ and $h_{max}$ lines that the swimmer cannot approach. If initially $\cos \psi <0$, then the swimmer will certainly reach the $h_{min}$ or $h_{max}$ line and the unboundedness of the orbit will depend on whether the point of interception with these lines is stationary and stable or not. {In the dynamical system that we consider here}, 	there is only one stationary point on $h_{min}$ and $h_{max}$ but it is unstable so the swimmer will  turn around and escape. 

As can be seen from Fig.~\ref{rhspace}A,  	the function $h_{max}(r)$ is monotonically increasing (this is easily verified analytically) but  $h_{min}$ has a maximum, which we find occurs analytically at the remarkably simple expression $r_m = [3(1-B)/2\alpha]^{1/2}$. 
Thus, the prolate swimmer will be trapped in a 3D rotlet vortex flow if and only if  both $r_0<r_m$ initially and $h(r_0,\psi_0)<h_{min}(r_m)$. Note that in the limit of spherical swimmers $B\to0$ we have $r_m = (3/2\alpha)^{1/2}$ which agrees with the result for a sphere in \S~\ref{sec::spherical}.

\subsection{Prolate spheroid in a 2D rotlet}\label{sec:prolate2D}
We now consider the case of a two-dimensional (2D) Stokes flow created by the rotation of an infinitely long cylinder of radius $R$ and angular velocity (along its axis) $\omega$, i.e.
\begin{equation}\bm{u} = \pmb{\omega}\times\bm{r}\frac{R^2}{r^2}\cdot
\end{equation} 
Interestingly, this flow has no vorticity so  spherical swimmers are not rotated by this flow. Thus, we focus on the prolate swimmers with shape factor $B>0$ for which the non-dimensional equations of motion are
\begin{subeqnarray}
	\dot{r} &=& \alpha \cos \psi, \\
	\dot{\phi} &=& \frac{\alpha}{r} \sin \psi +\frac{1}{r^2}, \\ 
	\dot{\psi} &=& -\frac{1}{r^2}(1+B\cos 2\psi)-\frac{\alpha}{r}\sin \psi, 
\end{subeqnarray}
with the same notation as above. Following a similar method to that for the 3D rotlet we obtain a conserved quantity
\begin{equation}
	h = \frac{\gamma K_0(z)+\sin \psi K_1(z)}{\gamma I_0(z)-\sin \psi I_1(z)},
\end{equation}
with $\beta=(1+B)/2$, $z = 2\sqrt{\beta x}=2\sqrt{\beta B}/\alpha r$ and $\gamma =  (\beta/B)^{1/2} = ((B+1)/2B)^{1/2}\geq1$.
Using the same method as in case of a 3D rotlet we come to a similar conclusion that a swimmer will be trapped in a 2D rotlet if and only if initially $r_0<r_m = (1-B)/\alpha$ and $h(r_0,\psi_0)<h_{min}(r_m)$. We compare in Fig.~\ref{rhspace}B this theoretical prediction with numerical computations, and again obtain excellent agreement.

\subsection{The trapping separatrix in the rod-like limit}

{In order to better understand the extent of trapping in vortical flows, we next investigate the boundary between the regions of bounded and unbounded orbits  in the $r-\psi$ space (i.e.~the separatrix).	This separatrix can only be implicitly defined as $h(r,\psi) = h(r_m,3\pi/2)$, since the bounded orbits are found in $r< r_m, h< h_{min}(r_m) = h(r_m,\psi = 3\pi/2)$ part of the $r-h$ space, as argued above. This implicit definition can be turned into an explicit one in  the limit of swimmers with large aspect ratios, $a\gg 1$. This rod-like limit is relevant not only for slender artificial microswimmers~\cite{ArtificialRodRheotaxis} but also for various types of bacteria~\cite{BOSO}.}  

Focusing on the 3D rotlet (the calculation in the 2D rotlet case is analogous), this implicit formulation takes the following form
\begin{equation}\label{eq:implicit}
	\frac{\gamma K_{-1/4}(z_m)- K_{3/4}(z_m)}{\gamma I_{-1/4}(z_m)+I_{3/4}(z_m)} = \frac{\gamma K_{-1/4}(z)+\sin \psi K_{3/4}(z)}{\gamma I_{-1/4}(z)-\sin \psi I_{3/4}(z)},
\end{equation}
where $z = 3\gamma B/2\alpha r^2$ and where $z_m = \gamma/2(\gamma^2-1)$ is the corresponding value of $z$ for $r = r_m$. Again, this equality is difficult to invert into an explicit formula $r(\psi)$;  however, in the limit of the swimmers that are very prolate ($B\to 1$ i.e.~$\gamma = [(B+1)/2B]^{1/2} \to 1$), 	it is possible to find an asymptotic expression for $z(\psi)$. In this limit, the maximal radius that the separatrix reaches is $r_m$  	so the minimal $z$ is $z_m \gg 1$. Thus we can assume $z\gg 1$ along the separatrix and we can use the well-known asymptotic expansion of the modified Bessel functions \cite{abramowitz1964handbook}
\begin{eqnarray}
	I_{\alpha}(z)&\sim&\frac{e^z}{\sqrt{2\pi z}}\left(1- \frac{4\alpha^2-1}{8z} +\frac{(4\alpha^2-1)(4\alpha^2-9)}{128 z^2}\right), \\	
	K_{\alpha}(z)&\sim&\sqrt{\frac{\pi}{2z}}e^{-z}\left(1+\frac{4\alpha^2-1}{8z} +\frac{(4\alpha^2-1)(4\alpha^2-9)}{128 z^2}\right),
\end{eqnarray} 
to find an approximate explicit expression for the separatrix. In the aforementioned limit, the expanded implicit formulation from Eq.~\eqref{eq:implicit} takes the   form
\begin{eqnarray}
	&e&^{-2(z-z_m)} = \frac{2\epsilon^2 + \mathcal{O}(\epsilon^3)}{2+\mathcal{O}(\epsilon)}\\
	&\times& \frac{1-\sin \psi+\epsilon+(3+ 5\sin \psi+3\epsilon)(32 z)^{-1}+\mathcal{O}(z^{-2})}{1+\sin \psi+\epsilon+(5\sin \psi-3-3\epsilon)(32 z)^{-1}+\mathcal{O}(z^{-2})},
\end{eqnarray}
with $\epsilon = \gamma - 1\ll 1$. 
\begin{figure}[t]
	\centering
	\includegraphics[width=0.65\linewidth]{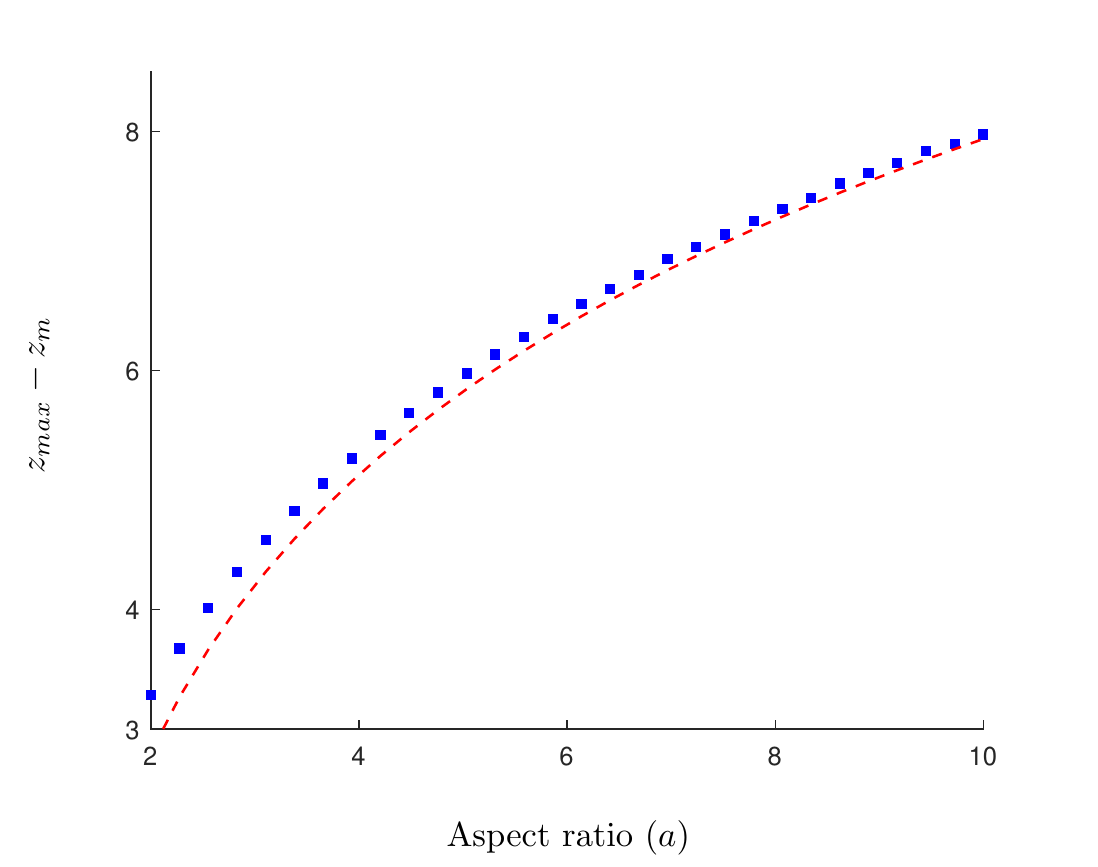}
	\caption{Numerical verification of the asymptotic expression for $z_{max}$ in a 3D rotlet flow. {Solid, blue squares show a numerical approximation obtained by inverting Eq.~\ref{eq:implicit} at $\psi = \pi/2$ for various values of the aspect ratio of the swimmers $a$. The red, dashed line follows the theoretical prediction based on Eq.~\ref{zmax}, while neglecting the $O(\epsilon)$ terms.}
	}\label{fig::numverzmax}
\end{figure}
As the balance of the dominant terms is different depending on $\sin \psi$, we further investigate each of these cases separately. In the $|\sin \psi \pm 1|\gg \epsilon$ case, the balance of dominant terms becomes
\begin{equation}
	\epsilon^2 \frac{1-\sin \psi}{1+\sin \psi} \approx e^{-2(z-z_m)},
\end{equation}
and thus we can approximate $z(\psi) = z_m + \ln (1/\epsilon) + \tanh^{-1} (\sin \psi)+\mathcal{O}(\epsilon)$. Now, if instead $\sin \psi = 1- \epsilon \zeta$, {with $\zeta\sim O(1)$}, the balance reads 
\begin{equation}
	\epsilon^2 \frac{(2+\zeta)\epsilon}{2} \approx e^{-2(z-z_m)},
\end{equation}
and the explicit formulation of the separatrix in this region is asymptotically $z = z_m + 1.5 \ln (1/\epsilon) - \ln \sqrt{1+\zeta/2}+\mathcal{O}(\epsilon)$. Finally, if $\sin \psi = -1+ \epsilon^2 \zeta$, with $\zeta\sim\mathcal{O}(1)$ we expect an $\mathcal{O}(1)$ correction of the form $z=z_m+F(\zeta)$ that must satisfy
\begin{equation}
	\frac{2}{2+\zeta+4F(\zeta)} \approx e^{-2F(\zeta)}.
\end{equation}
Solving this gives us $2F(\zeta) = -1-\zeta/2-W_{-1}[-\exp (-1-\zeta/2)]$, where $W_{-1}$ is a branch of the Lambert $W$ function.

An important result of this calculation is the value of the minimal radius that the separatrix reaches $r_{min}$, or equivalently, the value $z_{max}$ of the maximal $z$. It is attained for $\psi = \pi/2$, thus it can be approximated as
\begin{equation}
	z_{max} = z_m - 1.5 \ln \left(\frac{a}{\sqrt{a^2-1}}-1\right)+ \mathcal{O}(\epsilon). \label{zmax}
\end{equation} 
In Fig.~\ref{fig::numverzmax} we show the results of numerical computations for the  value of $z_{max}$ together with the theoretical predictions based on Eq.~\ref{zmax} for various values of the aspect ratio $a$ of the swimmer. 
We obtain excellent agreement across a large range of biologically relevant aspect ratios.

\section{Impact of noise on trapping \label{noise}}

The deterministic calculations in the previous sections provided insights into the behaviour of microswimmers in vortices. However,  the existence of bounded orbits seems to be at odds with the   depletion zones observed experimentally~\cite{depletion}. In order to  capture this experimentally observed phenomena, we  include in our model the effects of Brownian (translational and rotational) noise. 

\subsection{Mathematical modelling}

We  add to our  mathematical model delta-correlated white noise with rotational and translational diffusion coefficients denoted by $D_r$ and $D_t$, respectively; in addition to thermal noise, this allows us to also capture the effective impact of	variability in the translational velocity of biological swimmers. For simplicity, we focus here on swimmers in the flow due to a rotating sphere (i.e.~the 3D rotlet) in the rest of this section;  results follow similarly for the 2D rotlet case. 

The stochastic form of the dimensionless governing Eqs.~\eqref{eq::3dprolate1}-\eqref{eq::3dprolate3} now takes the following Langevin form    
\begin{subeqnarray}\slabel{neweqs1}
	dr &=& \underbrace{\alpha \cos \psi}_{\mu_r}\, dt\, + \sqrt{2} Pe_t^{-\frac{1}{2}} dW_t, \slabel{eq::stocR} \\ \slabel{neweqs2}
	d\psi &=& \underbrace{-\left[\frac{3}{2r^3}(1+B \cos 2\psi) +\frac{\alpha}{r}\sin \psi \right]}_{\mu_\psi} dt +\sqrt{2} Pe_r^{-\frac{1}{2}} dW_r, \hspace{5mm}\slabel{eq::stocPsi}
\end{subeqnarray}
where $W_t$ and $W_r$ are two independent Wiener processes, $Pe_t = \omega R^2/D_t$ is the translational P\'eclet number and $Pe_r = \omega/D_r$ is the rotational P\'eclet number. {The drifts $\mu_r$ and $\mu_\psi$ defined in these equations represent the deterministic rates of change of $r$ and $\psi$, respectively.} It should be noted that translational noise impacts the evolution of $\psi$ as well, since $\psi$ changes as the swimmer translates tangentially {to the circles of constant $r$}. 
For a {swimmer of typical size $d$}, it holds that $Pe_r/Pe_t = D_t/R^2 D_r = (d/R)^2\ll 1$~\cite{Stark2016} and hence we can neglect translational noise in the dynamics of $\psi$ relative to the rotational noise. { It is known that anisotropic particles, such as the prolate spheroids considered here,  experience anisotropic diffusion due to their different hydrodynamic mobilities along and perpendicular to its axis of symmetry~\cite{KimKarrila}. Since we only consider here a two-dimensional geometry (i.e.~the cells are restricted to move within a plane), this impacts  translational, but not    rotational,  diffusion. The anisotropy in mobility  means that, strictly speaking, the translational P\'eclet number is a function of the cell orientation, $\psi$ (this is a weak dependance since the maximum ratio in mobilities is known to be two, obtained for slender shapes). For simplicity, here we  assume that the P\'eclet number remains a constant, an assumption consistent with the fact that we  obtain no significant dependence of our results  on its exact value. }

\subsection{Quantifying the quality of trapping}
Since Eqs.~\eqref{neweqs1}-\eqref{neweqs2} are coupled, it is difficult to extract  statistical properties, such as mean square displacement. However, since noise is now included in the model, it is unlikely (i.e.~of probability zero) that swimmers starting on a bounded trajectory will remain trapped indefinitely. In order to quantify the quality of trapping, we thus focus our attention on finding the average time swimmers take to escape. 

This brings a few questions to consider. First, what does it mean exactly for a swimmer to escape? A natural choice would be to qualify the swimmers as escaped once they cross the (implicit) separatrix that delineates the regions of bounded and open orbits obtained above. However, as a swimmer gets further from the rotating body,  noise becomes dominant over the effects of the flow and thus we expect swimmers to return infinitely many times to the region of deterministically trapped orbits. Having the separatrix as the classifying boundary has therefore  limited physical significance. Instead, since there are  no deterministically trapped orbits with $r>r_m$, we set $r=r_m$ as the qualifying boundary. 

Another important point is what exactly happens once a swimmer reaches the rotating body located at $r=1$ (in dimensionless units)? The exact physics of the encounter between a microswimmer and a solid rotating body depends of course on the microscopic details of both the swimmer and the surface. In the specific case of bacteria {\it Bacillus subtilis} used in the experiments in Ref.~\cite{depletion}, the cells have been observed to stick to the surface of solid spheres. Thus, it is appropriate to  treat  $r=1$ as an absorbing boundary. 

For a swimmer that starts from $(r,\psi) = (r_0,\psi_0)$ at time $t=0$, we thus aim to determine the mean time $T(r_0,\psi_0)$ that it would take to reach either $r=r_m$ (escape from the vortex from the outside) or $r=1$ (reaching the surface of the rotating body). While the mean first passage time (MFPT) $T(r_0,\psi_0)$ is a useful measure for quantifying the quality of trapping, it does not allow us to distinguish between the two types of escapes (away from the body vs.~sticking to it). Thus, we will also introduce the probability $P(r_0,\psi_0)$ the a swimmer will reach $r=r_m$ (and thus escapes) before hitting the rotating body, as the second important statistic of the problem.

\subsection{Fokker--Planck formalism}

A common approach to solving such drift-diffusion problems is to introduce the Fokker--Planck equation 
\begin{equation}
	\frac{\partial p}{\partial t} = -\frac{\partial }{\partial r} p\mu_r -\frac{\partial }{\partial \psi} p\mu_\psi + \left[Pe_t^{-1} \frac{\partial^2 }{\partial r^2} + Pe_r^{-1} \frac{\partial^2 }{\partial \psi^2}\right]p \triangleq \mathcal{L} p, \label{eq::FP}
\end{equation}
that governs the evolution of the probability distribution $p(r,\psi,t|r_0,\psi_0)$ for  a swimmer located at $(r,\psi)$ at time $t$ starting from $(r_0,\psi_0)$ at $t=0$ (e.g. see Ref.~\cite{FP_book}). To determine statistics of the system, such as the MFPT or the escape probability, we need to introduce the adjoint of the Fokker--Planck equation \cite{mfpt_paper,mfpt_book}. Using $\mathcal{L}$ to denote the Fokker--Planck operator  from Eq.~\eqref{eq::FP},  its adjoint $\mathcal{L}^\dagger$ is defined  on the space of the initial conditions $(r_0,\psi_0)$ {as the unique operator satisfying 
	\begin{equation}
		\langle\mathcal{L} f, g \rangle = \langle f,\mathcal{L}^\dagger g \rangle,
	\end{equation}
	for any two functions $f$ and $g$, where $\langle \cdot,\cdot \rangle$ denotes the standard integral inner product defined as
	\begin{equation}
		\langle f,g \rangle = \int_{0}^{2\pi} \int_1^{r_m} f^* g\, dr d\psi,
	\end{equation}
	where $f^*$ is the complex conjugate of the function $f$.} Following standard algebra, 	we thus obtain $\mathcal{L}^\dagger$ as
\begin{equation}
	\mathcal{L}^\dagger = Pe_t^{-1} \frac{\partial^2}{\partial r_0^2} + Pe_r^{-1}  \frac{\partial^2}{\partial {\psi_0}^2} +\mu_r \frac{\partial}{\partial {r_0}} + \mu_\psi \frac{\partial}{\partial {\psi_0}}.
\end{equation}

\begin{figure*}
	\centering
	\includegraphics[width=\linewidth]{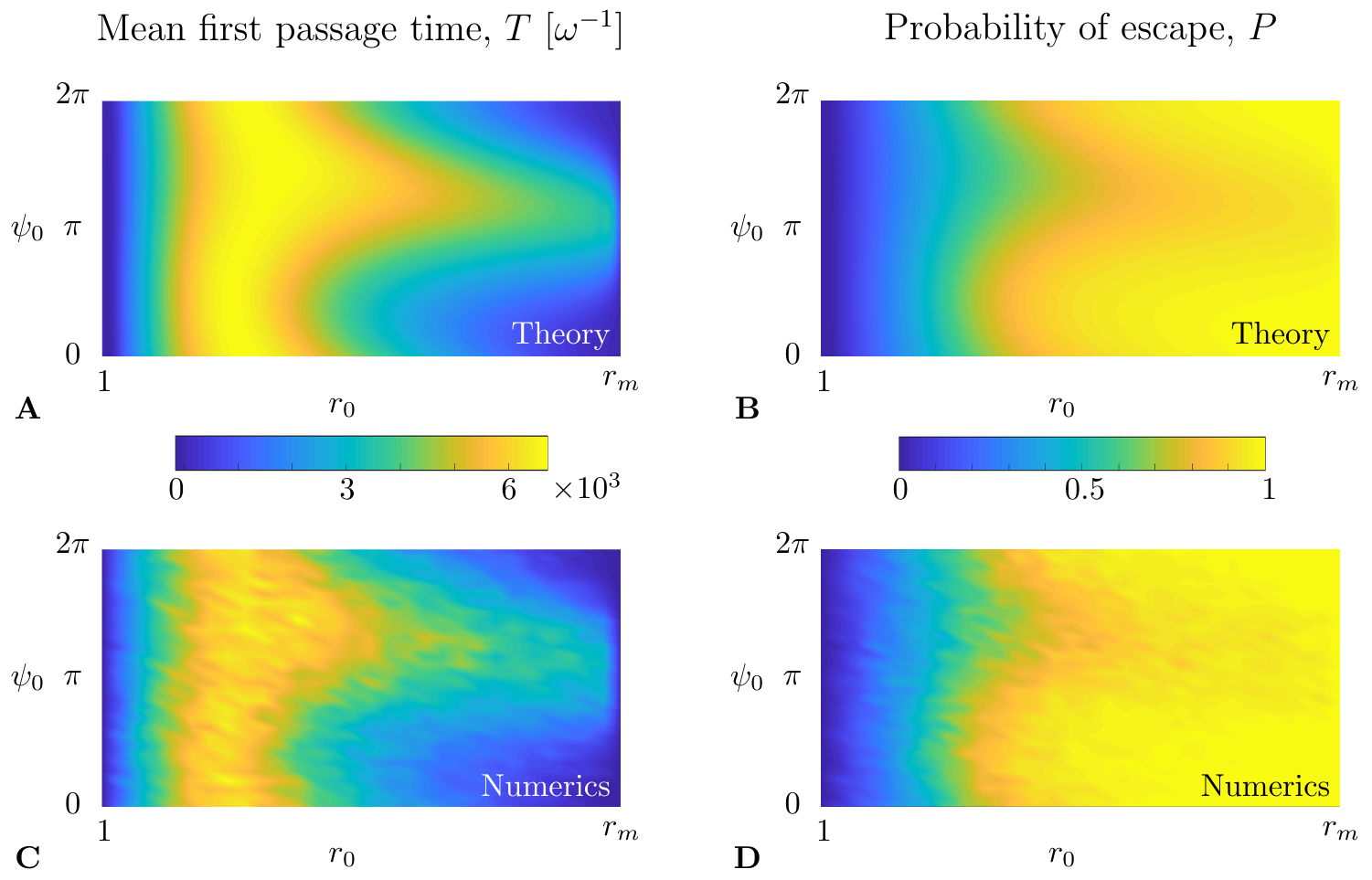}
	\caption{Comparison between  theory and simulations for the mean first passage time $T(r_0,\psi_0)$ (A and C, in units of $\omega^{-1}$) 
		and probability of escape $P(r_0,\psi_0)$ (B and D). 
		A and B are obtained by direct integration of the adjoint Fokker--Planck equations, Eq.~\eqref{eq::mfpt}-\eqref{eq::prob}, while C and D are results of direct numerical simulations of the Langevin swimming model in Eqs.~\eqref{eq::stocR}-\eqref{eq::stocPsi}. Results shown are for  $Pe_t^{-1} = 0.045, Pe_r^{-1} = 0.05, B=0$ and $\alpha = 0.008$.}\label{MFPTfig}
\end{figure*}

Using the adjoint operator, it is a classical results that the mean first passage time $T$ and the probability of escape are known to satisfy the  partial differential equations~\cite{FP_book}
\begin{eqnarray}\label{FP1}
	\mathcal{L}^{\dagger} T &= -1 \hskip 3pt  &\text{with} \hskip 3pt   T=0 \hskip 3pt  \text{at} \hskip 3pt  r_0=1  \hskip 3pt \text{and}  \hskip 3pt r_0 = r_m ,\label{eq::mfpt} \\
	\mathcal{L}^{\dagger} P &= 0 \hskip 3pt   &\text{with} \hskip 3pt   P = 0 \hskip 3pt  \text{at} \hskip 3pt  r_0 = 1 \hskip 3pt   \text{and} \hskip 3pt P = 1 \hskip 3pt \text{at} \hskip 3pt  r_0 = r_m. \label{eq::prob}
\end{eqnarray}

\subsection{Solution of Fokker-Planck model and validation}	

We solve  Eq.~\eqref{eq::mfpt}-\eqref{eq::prob} numerically   using  the open-source  \verb|FreeFem++| 	\cite{freefem} solver, based on the finite element method. The results for the mean first passage time
$T(r_0,\psi_0)$ { are shown in} Fig.~\ref{MFPTfig}A (in units of $\omega^{-1}$) while the  probability of escape 
$P(r_0,\psi_0)$ is plotted in Fig.~\ref{MFPTfig}B. 
To verify the results of our numerical integration,  we also compare it with  ensemble averages of a large number of realisations of direct numerical time-stepping of the Langevin model  in Eqs.~\eqref{eq::stocR}-\eqref{eq::stocPsi} using the Runge--Kutta fourth-order method. These results are plotted in Fig.~\ref{MFPTfig}C-D below the corresponding Fokker-Planck result, where we obtain excellent {quantitative} agreement. 

Apart from the obvious observations that MFPT is small next to the boundaries and larger in between, we observe that the region of high MFPT extends significantly close to the $r=r_m$ boundary along a $\psi\sim \pi$ direction (see Fig.~\ref{MFPTfig}A or C). Intuitively, this can be explained by noting that $\psi\sim\pi$ means that the swimmer is directed towards the decreasing $r$ direction; a swimmer  close to  $r=r_m$ has therefore a high chance of escaping through that boundary (see Fig.~\ref{MFPTfig}B or D) and  will take a long time to turn around and escape. 

Following the same logic, one would expect a region of low MFPT to extend from the $r=1$ boundary along the $\psi\sim \pi$ direction. However, in that region, the flow vorticity is strong and rotates the swimmers rapidly; the radial swimming vanishes thus on average there and  escape is instead dominated by   translational diffusion, which is independent of the swimmers' orientation. Based on the same arguments, we can expect the probability of escape to be lower for $\psi\sim \pi$ in the region close to $r=r_m$ and independent of $\psi$ close to $r=1$, in agreement with results  in Figs.~\ref{MFPTfig}B and D.

\subsection{Parameter sweep}	\label{sec:sweep}
Next, we explore the phase space of the four dimensionless parameters of the problem: (i) the relative swimming speed $\alpha$, (ii) the swimmer's shape factor $B$ and the two P\'eclet numbers (iii) $Pe_t$ and (iv) $Pe_r$. In order to quantify the overall quality of trapping as a function of the parameters, we consider some scalar moments of the continuum solutions for the MFPT $T(r_0,\psi_0)$ and the probability of escape $P(r_0,\psi_0)$.
A natural first choice is the average probability of escaping $\langle P \rangle$, taken over the range $r\in [1,r_m]$. Note that we want the swimmers to initially be uniformly distributed in space i.e.~according to a linear distribution in $r$, $f_U(r_0,\psi_0) = r_0/\pi (r_m^2-1)$ so that 
\begin{equation}
	\langle P \rangle = \int_{0}^{2\pi}\int_1^{r_m} \frac{P(r_0,\psi_0)r_0}{\pi (r_m^2-1)}\,dr_0\,d\psi_0. \label{eq:avgp}
\end{equation}
Additionally, we measure the maximal MFPT, denoted by $\tilde{T}$, which  sets a timescale for the emergent dynamics of the ensemble of swimmers.

\begin{figure*}
	\centering
	\includegraphics[width=.8\linewidth]{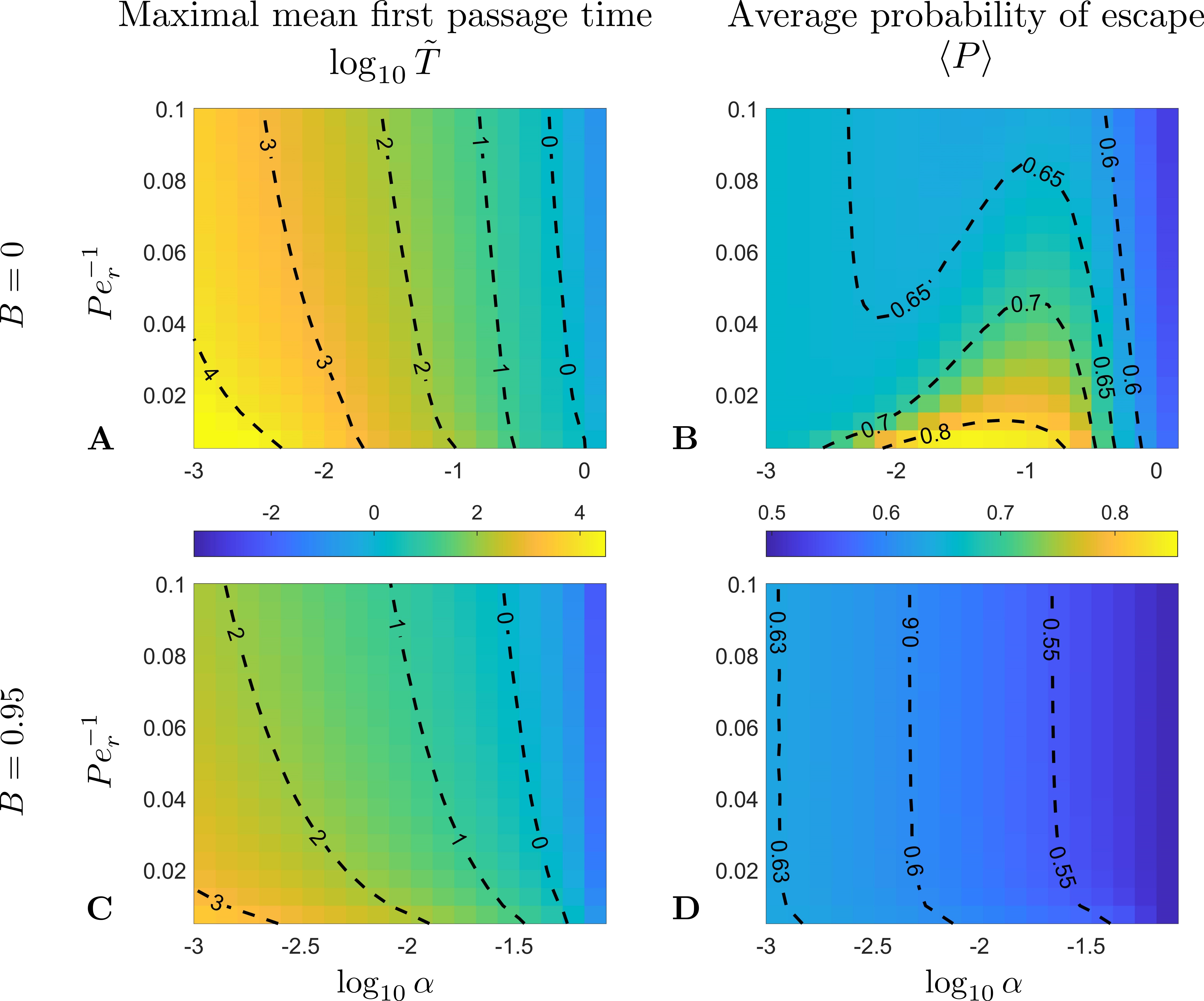}
	\caption{(A,C) Maximal MFPT, $\tilde{T}$, and average probability of escape, $\langle P \rangle$, for two different values of the shape parameter $B=0$ (spherical limit; A,B) and $B=0.95$ (elongated swimmer; C,D), as a function of the dimensionless swimming speed, $\alpha$, and the rotational P\'eclet number, $Pe_r$. 
	}
	\label{fig::sweep}
\end{figure*}

In computations, we   fix the ratio between the translational P\'eclet number and the  rotational one to $Pe_t/Pe_r = 4 $; the exact value for this ratio turns out to not affect the qualitative features of our results, as confirmed  in our computations (not shown). We then explore   the rest of the parameter space by fixing the shape parameter $B$  and then sweeping among the values for the pair $(\alpha, Pe_r)$. In Fig.~\ref{fig::sweep} we show results for two limiting values of $B$: spherical swimmers with $B=0$ (Figs.~\ref{fig::sweep}A and B) and  rod-like swimmers with $B=0.95$ (Figs.~\ref{fig::sweep}C and D).

\begin{figure*}
	\centering
	\includegraphics[width=0.81\linewidth]{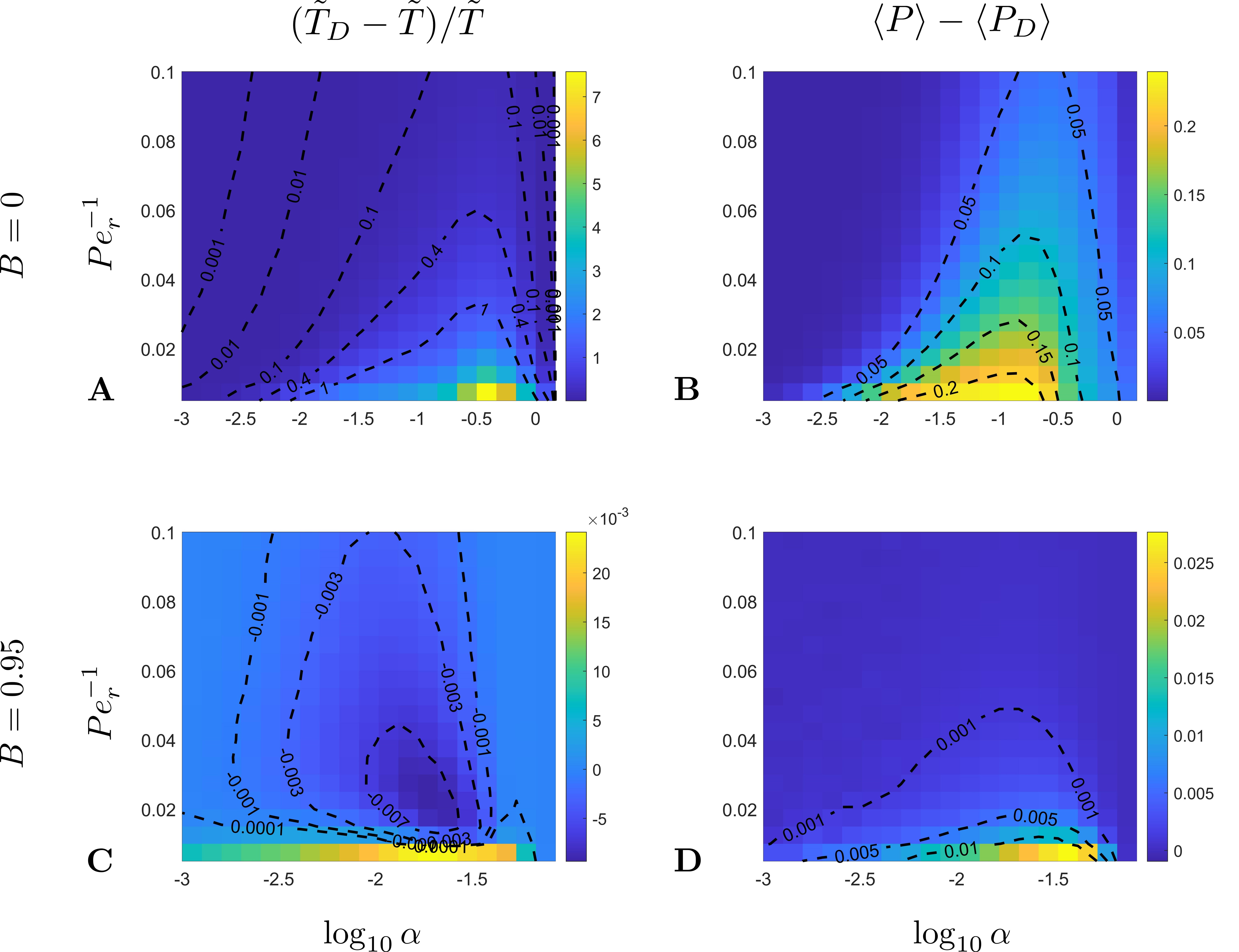}
	\caption{Comparison between numerical results (from  Fig.~\ref{fig::sweep}) for the maximal MFPT, $\tilde{T}$ (A,C), and the average probability of escape, $\langle P \rangle$ (B,D),   and the theoretical predictions   in the diffusion dominated regime, $\tilde{T}_D$ and $\langle P_D \rangle$, Eq.~\eqref{DDregime}.   (A,B): spherical swimmers $B=0$;     (C,D):   rod-like swimmers $B=0.95$.}
	\label{fig::sweep_diff}
\end{figure*}
The maximal MFPT ($\tilde{T}$) behaves qualitatively as expected. The faster the swimmers  and the stronger the noise, the easier it is to escape and thus $\tilde{T}$ takes smaller values. The difference in maximal MFPT between the two values of the shape factor $B$  can be understood by realising that the trapping domain is different for distinct values of $B$ and $\alpha$. Indeed, the trapping domain is given by $r\in[1,r_m]$ where $r_m = [3(1-B)/2\alpha]^{1/2}$ so it is significantly smaller for rod-like swimmers ($B\to 1$) than   for   spherical ones ($B=0$). The size of the trapping region indicates that  more elongated swimmers are more difficult to trap, which is   reflected in the fact that MFPT is shorter for elongated swimmers.

Regarding the  average probability of escape, we notice in  Figs.~\ref{fig::sweep}B and D   that for strong noise (large $Pe_r^{-1}$) the value of $\langle P\rangle$ is decreasing with $\alpha$. This is also due to the fact that the decrease in $\alpha$ leads to an increase in $r_m$ i.e.~the size of the trapping domain. Recall that the swimmers are initially distributed uniformly in space, so they follow a linear distribution in $f_U\sim r_0$ and the average probability of escape is   weighted  towards larger values of the initial position $r_0$ (see Eq.~\ref{eq:avgp}). With strong noise, the increase in $r_m$ leads to even more weight towards  $r_0=r_m$ and a larger escape probability (this is also captured by Eq.~\eqref{PD} below).

Interestingly, we  note that in Figs.~\ref{fig::sweep}B and D, the value of $\langle P \rangle$ appears to always be in the range $ 0.6-0.7$ for small values of the non-dimensional swimming speed $\alpha$. 
To rationalise this observation, we can use an analysis  in the diffusion-dominated regime. 	When locomotion is negligible  (i.e.~$\alpha \ll 1$),   the adjoint system of equations, Eqs.~\ref{eq::mfpt} and \ref{eq::prob}, admits simple solutions given by
\begin{subeqnarray}
	T_D &=& \frac{Pe_t}{2} (r-1)(r_m-r), \\
	P_D &=& \frac{r-1}{r_m-1},
\end{subeqnarray}
with the subscript $D$ used to   indicate this is the diffusion-dominated solution. This in turn allows to  compute  the two trapping measures as
\begin{subeqnarray}\label{DDregime}
	\tilde{T}_D &=& \frac{Pe_t}{8}(r_m-1)^2, \\
	\langle P_D \rangle &=& \frac{2r_m+1}{3r_m+3}.\slabel{PD}
\end{subeqnarray}
Since $r_m = [3(1-B)/2\alpha]^{1/2}$,    small values of $\alpha$ correspond to $r_m \gg 1$, leading to the approximate solution $\langle P_D \rangle \approx 2/3$. The small-$\alpha$ limits in Figs.~\ref{fig::sweep}B and D correspond therefore a constant probability 
of $2/3 $.

In Fig.~\ref{fig::sweep_diff} we further compare the computational results from Fig.~\ref{fig::sweep} with the diffusion-dominated predictions in Eq.~\eqref{DDregime}. The dark blue regions are those for which the simulations are in line with the no-swimming predictions, and we observe a systematic agreement for small values of $\alpha$. We also note a significant deviation from these predictions around the $\alpha\sim 0.1, Pe_r^{-1}\sim 0.01$ region where the maximal MFPT can become up to 10 times smaller than the pure diffusion value. It is also notable that it takes a much stronger noise to overcome swimming for spherical swimmers (Figs.~\ref{fig::sweep_diff}A,B) than it does for rod-like swimmers (Figs.~\ref{fig::sweep_diff}C,D).  	Here again  this is due to the fact that the trapping region is smaller for more elongated particles.

Importantly, the results in Fig.~\ref{fig::sweep_diff} indicate that the region where the maximal MFPT $\tilde{T}$ is significantly less than $\tilde{T}_D$  coincides roughly with the region where the average probability of escape is significantly greater than the diffusion dominated prediction, a result that appears to hold for both limiting values of the shape factor $B$. Consequently, swimming not only promotes faster migration from the trapping region but it does  so by increasing the chance of escape and not by making more particles stick to the rotating body.

\begin{figure*}
	\centering
	\includegraphics[width = .6\linewidth]{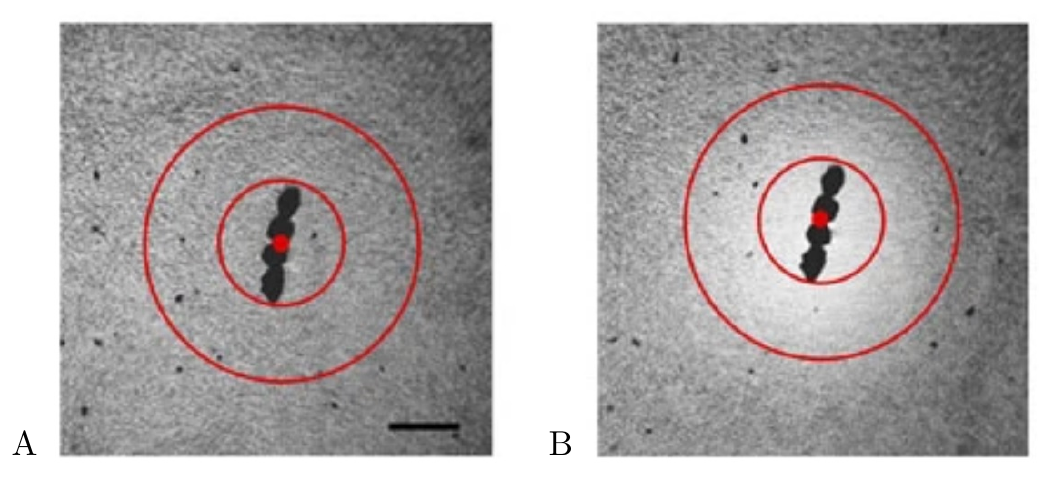}
	\caption{Formation of a depletion zone of swimming bacteria around a rotating sphere. A: Initial distribution of bacteria. B: Depletion zone formed (lighter colour, lower concentration) after 50~s of rotation at 5~Hz. Adapted with permission from Ref.~\cite{depletion}, licenced under \href{https://creativecommons.org/licenses/by/4.0/}{CC BY 4.0}.}\label{fig::deplorg}	
\end{figure*}

\section{Depletion zone and comparison with experiments \label{sec:depletion}}

In this section, we show how the theoretical framework developed above can explain the formation of the depletion of swimming bacteria around a rotating sphere reported in Ref.~\cite{depletion}; these  experiments are reproduced in Fig.~\ref{fig::deplorg}, with the initial distribution of bacteria shown in Fig.~\ref{fig::deplorg}A and the depletion (shown in a lighter colour, corresponding to a lower concentration) visualised in Fig.~\ref{fig::deplorg}B. 

The original study included a theoretical model, with bacteria modelled as self-propelled rods ($B=1$) swimming in a 2D rotlet flow~\cite{depletion}. The comparison with the experiments was then done by performing the numerical integration of the Fokker-Plank equations with a reflecting boundary condition at the surface of the rotating object. 
{ Although the agreement between the theoretical predictions from Ref.~\cite{depletion} and their experimental results  appear good, an assumption of their model  disagrees qualitatively with observations. It was noted that, experimentally, about 40\% of bacteria initially located in what becomes  the depletion zone end up stuck to the surface of the rotating body; this clearly contradicts the assumption that the surface of the body is reflective. 
	
	Thus, to predict the experimental results,  we propose here that the more appropriate   boundary condition at the surface of the rotating body is for it to be absorbing. We also consider a more realistic value of the shape factor ($B<1$); the rest of the model is similar to the one in Ref.~\cite{depletion}. Instead of solving the Fokker-Plank equations, we   estimate the concentration profile by using a Monte Carlo method based on the agent based simulations of the corresponding evolution equations.}

{ To confirm that altering the boundary condition from reflecting to absorbing does not prevent the depletion zone from forming,} we start by carrying out direct swimming simulations on an ensemble of prolate swimmers by numerically time-stepping   Eqs.~\eqref{eq::stocR}-\eqref{eq::stocPsi}. As a test case, we choose a frequency of rotation of $f=20$Hz, cells of shape factor $B=0.85$ (corresponding to aspect ratio $a \approx 3.51$)
and take {the values for the average swimming speed and diffusion coefficients} from the data in Ref.~\cite{depletion}. { Note that the particular value of $B$ was chosen to reflect the aspect ratio of bacteria used in Ref.~\cite{depletion}. However, additional simulations (not shown) reveal that that the results are not particularly sensitive to the choice of $B$.}	We initialise a uniform distribution of  3112 non-interacting swimmers in the region $1\leq r \leq L$, where $L\gg R_d$ with $R_d$ being  the radius of the depletion zone  reported experimentally for $f=20$Hz. Here the  boundary at $r=1$ (i.e.~the rotating sphere) is set to be absorbing while $r=L$ is taken to be reflective. Note that a spatially uniform distribution of bacteria implies a linear distribution in $r$ in the $(r,\psi)$ space, i.e.~a uniform distribution $f_U(r,\psi) = r \pi^{-1}(L^2-1)^{-1}$.

\begin{figure*}
	\centering
	\includegraphics[width=\linewidth]{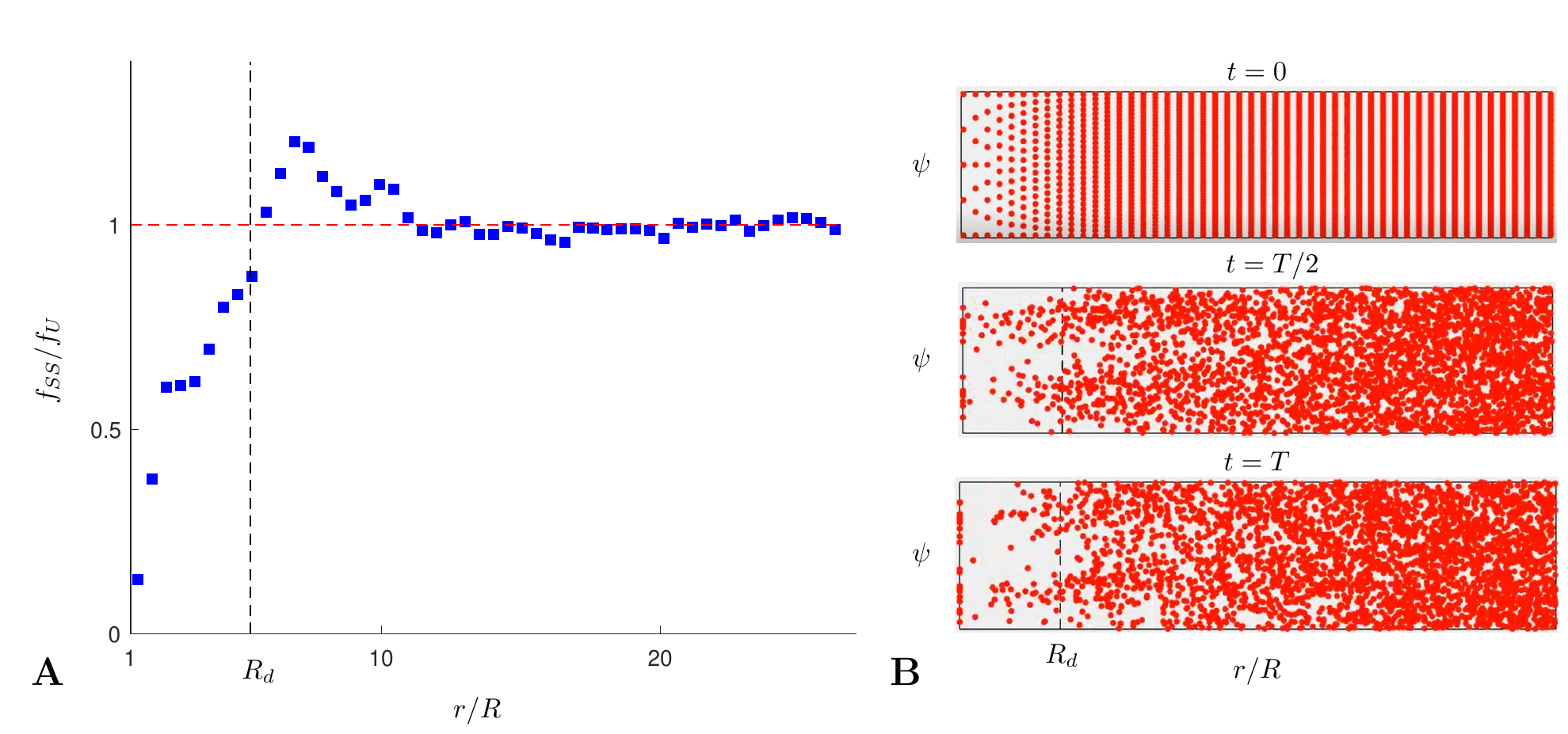}
	\caption{Numerical verification of the emergence of the depletion zone in our theoretical model. 
		A: Steady-state distribution $f_{SS}$ of an ensemble of non-interacting swimmers following the stochastic model in Eqs.~\eqref{eq::stocR}-\eqref{eq::stocPsi}, normalized to the uniform distribution and averaged over the angle $\psi$. 
		B: Snapshots from the  simulations used to obtain data for A; shown at the start ($t=0$), midway through ($t=T/2$) and at the end of the simulation ($t=T$). The ensemble contains 3112 swimmers and the total simulation time is $T=8000$, {in   units of $\omega^{-1}$}.}\label{fig::depletionfig}
\end{figure*}

{We carry out these computations until a steady distribution of swimmers is obtained, which in our simulations happens before $T=8000$ (in  units of $\omega^{-1}$),  so this is when we stop our computations. The resulting distribution, averaged over the angle $\psi$, and denoted by $f_{SS}$,  is shown in Fig.~\ref{fig::depletionfig}A relative to a uniform distribution $f_U$. We see the clear emergence of a depletion region at $r<R_d$
	where the concentration of cells is significantly less than its initial value (recall that $R_d$ is the radius of the depletion zone  from experiments). In Fig.~\ref{fig::depletionfig}B we show snapshots of the particular realisation of the ensemble simulations at times $t=0, T/2$, $T$; here again the depletion zone is apparent at $t=T$.} Since all  the parameters were taken from the experiments, including the   depletion radius $R_d$, we conclude that our Langevin approach correctly captures the steady-state depletion of swimming bacteria in the experiments.

{ Using additional simulations (not shown), we have also confirmed the emergence of the depletion zone for various values of the probability $p_{st}$ of the swimmer sticking to the rotating body after each encounter, from $p_{st} = 0$ (perfectly reflective surface) to $p_{st}=1$ (perfectly absorbing surface). We have found that the size of the depletion zone varies weakly with the value of $p_{st}$. This result is consistent  wirh the model in Ref.~\cite{depletion} that assumed a reflective surface and still recovered the emergence of the depletion. The escape away from the rotating body is therefore the critical component in the establishment of the depletion zone. }

With the emergence of the depletion zone confirmed quantitatively in the model, we can now  compare our theoretical predictions for $R_d$ with the experimental measurements. To  estimate the value of $R_d$, we use the experimental observation that roughly 40\% of bacteria from the depletion zone get stuck at the rotating body. Thus, we can use the probability of escape $P(r,\psi)$ described  above to define the depletion zone radius $R_d$ as the solution  
\begin{equation} 
	\int_{1}^{R_d} \int_{0}^{2\pi} \frac{r P(r,\psi)}{\pi (R_d^2 - 1)} \, d\psi \, dr = 0.6, \label{eq::rd_def}
\end{equation}
{i.e.~as the radius $R_d$ such that the swimmers that start closer than $R_d$ to the rotating body have a 60\% probability of escape, or equivalently, 40\% probability of getting stuck to the rotating body.}
With $B=0.85$ 	and taking all parameters from the  experiments, we compare in 
Table~\ref{tb::depl_radius} our prediction for the depletion radius $R_d$  (in units of the sphere radius $R$) with the experimentally measured values in Ref.~\cite{depletion} for four rotation frequencies of the sphere (from 10~Hz to 40~Hz).  
We obtain excellent quantitative agreement between the experiments and our Fokker--Planck continuum predictions in all cases. 		

\begin{table}[b]
	\small
	\begin{tabular*}{0.48\textwidth}{@{\extracolsep{\fill}}lll}
		\hline
		Bead frequency [Hz] & Experiments Ref.~\cite{depletion} & Theory \\
		\hline
		3.0 & 2.5 $\pm$ 0.3 & 2.0 \\
		10 & 4.1 $\pm$ 0.4 & 3.6 \\
		20 & 5.1 $\pm$ 0.5 & 5.0 \\
		40 &  6.5 $\pm$ 0.5 & 6.9 \\
		\hline
	\end{tabular*}
	\caption{\label{tb::depl_radius} Comparison of the experimental~\cite{depletion} and theoretical results for the radius of the depletion zone { $R_d$} in the units of the bead radius {$R$}. Theoretical predictions are calculated under the assumption $B=0.85$ (i.e.~for a cell with  effective aspect ratio $a=3.51$).}
\end{table}

\begin{figure}
	\centering
	\includegraphics[width = .9\linewidth]{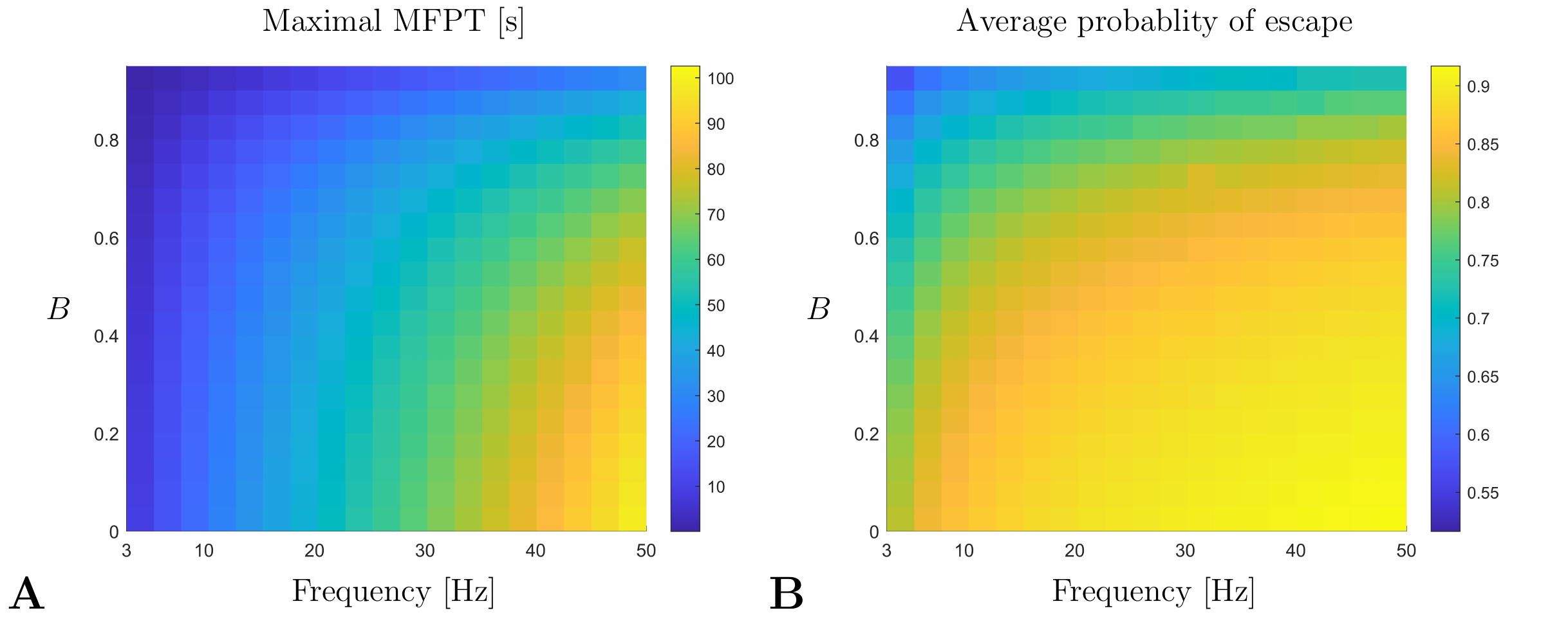}
	\caption{Maximal MFPT (A) and   average probability of escape (B) as a function of the rotation frequency of the sphere, $\omega$, and of the shape factor of the swimmer $B$. All  parameters of the problem were fixed according to experimental data from Ref.~\cite{depletion} and the MFPT is given in units of seconds [s]. }\label{fig::tmax}	
\end{figure}

{ Finally, by using the continuum model computations of the first passage time and the probability of escape, as described in the previous section, we may investigate how the quality of trapping changes with the dimensional parameters of the experiment, namely the frequency of rotation $\omega$ and the shape factor of the swimmers $B$. As opposed to the parameter sweep done in \S\ref{sec:sweep}, we  vary here only these two parameters since: (i) the frequency $\omega$ is the only directly controllable parameter of the problem and (ii) $B$ is an effective shape parameter that is not easily measured for a population of microorganisms.  
	As argued above, the quality of trapping is well described by the maximal MFPT and the average probability of escape.} In Fig.~\ref{fig::tmax} we plot the relevant results as derived from the data obtained by numerically solving the Fokker-Planck model in Eqs.~\eqref{eq::mfpt}--\eqref{eq::prob}. The quality of trapping is seen to increase with the frequency of rotation, with the vortex flow able to overpower both  swimming and noise. However, trapping becomes more difficult  as the   swimmers become more elongated (i.e.~when $B$ increases); this is due to persistence in swimming orientation for elongated swimmers, which promotes a faster escape. Note that the typical time-scale of the maximal MFPT is predicted to be around 50~s, a value which agrees with the observed time of formation of the depletion zone~\cite{depletion}.

\section{Conclusions \label{sec::conclusion}}

In this paper, we investigate the behaviour of microswimmers in elementary vortices, namely in two- and three-dimensional   rotlet flows. In the absence of noise, a  mathematical approach reveals the existence of deterministically bounded orbits near the centre of the vortex and unbounded ones further away. For elongated microswimmers, we discover a conserved quantity of motion that allows us to easily map the regions of phase space according to the type of the orbit (bounded/unbounded). Next, we introduce translational and rotational noise into the system, modelling both thermal  noise as well as active biological fluctuations. We quantify the quality of trapping in the deterministically bounded orbits near the centre of the vortex by examining the probability of escape and mean escape time of the swimmers starting on  said deterministically bounded orbits. We show how to use these findings to formulate a prediction for the radius of the depletion zone (see Fig.~\ref{fig::deplorg}), which compares favourably with the experiments~\cite{depletion}. 

In the case of spherical swimmers, the  equations in our model were found to have a Hamiltonian structure. The axial symmetry of the flow also leads to a conserved quantity of motion for elongated swimmers, even though the system is no longer Hamiltonian. Interestingly, a similar behaviour was reported for swimmers in   Poiseuille flow~\cite{zottl2013periodic}, suggesting an underlying fundamental behaviour  in  the two-dimensional swimming of elongated microswimmers in Stokes flows.

Throughout this paper, we have assumed the motion to remain two-dimensional. This simplification is supported by the experimental setup in Ref.~\cite{depletion} in which bacteria are hydrodynamically attracted to, and thus gather on, a horizontal  surface below the rotating sphere.  	Allowing swimmers to explore the third dimension in the model of the three-dimensional rotlet, or performing similar experiments but in bulk fluid, could { potentially} lead to new behaviour not seen within the two-dimensional assumption. In this paper, we demonstrated that the swimmers that are relatively close to the rotating body would be deterministically trapped in the plane orthogonal to the axis of rotation of the body. Allowing motion in the third dimension { would represent an important extension of the current work as it} could result in plumes of swimmers escaping along the axis of rotation, and therefore an interesting new instance of collective motion.

\section*{Acknowledgements}
This project has received funding from the European Research Council under the European Union's Horizon 2020 research and innovation program (Grant No. 682754 to E.L.). It was also funded by Trinity College, Cambridge (IGS scholarship to	I.T.).	
\newpage


\begin{thebibliography}{57}%
	\makeatletter
	\providecommand \@ifxundefined [1]{%
		\@ifx{#1\undefined}
	}%
	\providecommand \@ifnum [1]{%
		\ifnum #1\expandafter \@firstoftwo
		\else \expandafter \@secondoftwo
		\fi
	}%
	\providecommand \@ifx [1]{%
		\ifx #1\expandafter \@firstoftwo
		\else \expandafter \@secondoftwo
		\fi
	}%
	\providecommand \natexlab [1]{#1}%
	\providecommand \enquote  [1]{``#1''}%
	\providecommand \bibnamefont  [1]{#1}%
	\providecommand \bibfnamefont [1]{#1}%
	\providecommand \citenamefont [1]{#1}%
	\providecommand \href@noop [0]{\@secondoftwo}%
	\providecommand \href [0]{\begingroup \@sanitize@url \@href}%
	\providecommand \@href[1]{\@@startlink{#1}\@@href}%
	\providecommand \@@href[1]{\endgroup#1\@@endlink}%
	\providecommand \@sanitize@url [0]{\catcode `\\12\catcode `\$12\catcode
		`\&12\catcode `\#12\catcode `\^12\catcode `\_12\catcode `\%12\relax}%
	\providecommand \@@startlink[1]{}%
	\providecommand \@@endlink[0]{}%
	\providecommand \url  [0]{\begingroup\@sanitize@url \@url }%
	\providecommand \@url [1]{\endgroup\@href {#1}{\urlprefix }}%
	\providecommand \urlprefix  [0]{URL }%
	\providecommand \Eprint [0]{\href }%
	\providecommand \doibase [0]{https://doi.org/}%
	\providecommand \selectlanguage [0]{\@gobble}%
	\providecommand \bibinfo  [0]{\@secondoftwo}%
	\providecommand \bibfield  [0]{\@secondoftwo}%
	\providecommand \translation [1]{[#1]}%
	\providecommand \BibitemOpen [0]{}%
	\providecommand \bibitemStop [0]{}%
	\providecommand \bibitemNoStop [0]{.\EOS\space}%
	\providecommand \EOS [0]{\spacefactor3000\relax}%
	\providecommand \BibitemShut  [1]{\csname bibitem#1\endcsname}%
	\let\auto@bib@innerbib\@empty
	\bibitem [{\citenamefont {Lighthill}(1975)}]{lighthill75}%
	\BibitemOpen
	\bibfield  {author} {\bibinfo {author} {\bibfnamefont {J.}~\bibnamefont
			{Lighthill}},\ }\href@noop {} {\emph {\bibinfo {title} {Mathematical
				Biofluiddynamics}}}\ (\bibinfo  {publisher} {SIAM},\ \bibinfo {address}
	{Philadelphia},\ \bibinfo {year} {1975})\BibitemShut {NoStop}%
	\bibitem [{\citenamefont {Lauga}(2020)}]{lauga_book}%
	\BibitemOpen
	\bibfield  {author} {\bibinfo {author} {\bibfnamefont {E.}~\bibnamefont
			{Lauga}},\ }\href@noop {} {\emph {\bibinfo {title} {The Fluid Dynamics of
				Cell Motility}}}\ (\bibinfo  {publisher} {Cambridge University Press},\
	\bibinfo {address} {UK},\ \bibinfo {year} {2020})\BibitemShut {NoStop}%
	\bibitem [{\citenamefont {Wheeler}\ \emph {et~al.}(2019)\citenamefont
		{Wheeler}, \citenamefont {Secchi}, \citenamefont {Rusconi},\ and\
		\citenamefont {Stocker}}]{StokerReviewBacteriaPlankton}%
	\BibitemOpen
	\bibfield  {author} {\bibinfo {author} {\bibfnamefont {J.~D.}\ \bibnamefont
			{Wheeler}}, \bibinfo {author} {\bibfnamefont {E.}~\bibnamefont {Secchi}},
		\bibinfo {author} {\bibfnamefont {R.}~\bibnamefont {Rusconi}},\ and\ \bibinfo
		{author} {\bibfnamefont {R.}~\bibnamefont {Stocker}},\ }\bibfield  {title}
	{\bibinfo {title} {Not just going with the flow: The effects of fluid flow on
			bacteria and plankton},\ }\href
	{https://doi.org/10.1146/annurev-cellbio-100818-125119} {\bibfield  {journal}
		{\bibinfo  {journal} {Annu. Rev. Cell Dev. Bi.}\ }\textbf {\bibinfo {volume}
			{35}},\ \bibinfo {pages} {213} (\bibinfo {year} {2019})},\ \bibinfo {note}
	{pMID: 31412210}\BibitemShut {NoStop}%
	\bibitem [{\citenamefont {Fauci}\ and\ \citenamefont
		{Dillon}(2006{\natexlab{a}})}]{fauci06}%
	\BibitemOpen
	\bibfield  {author} {\bibinfo {author} {\bibfnamefont {L.~J.}\ \bibnamefont
			{Fauci}}\ and\ \bibinfo {author} {\bibfnamefont {R.}~\bibnamefont {Dillon}},\
	}\bibfield  {title} {\bibinfo {title} {Biofluidmechanics of reproduction},\
	}\href@noop {} {\bibfield  {journal} {\bibinfo  {journal} {Annu. Rev. Fluid
				Mech.}\ }\textbf {\bibinfo {volume} {38}},\ \bibinfo {pages} {371} (\bibinfo
		{year} {2006}{\natexlab{a}})}\BibitemShut {NoStop}%
	\bibitem [{\citenamefont {Guasto}\ \emph {et~al.}(2012)\citenamefont {Guasto},
		\citenamefont {Rusconi},\ and\ \citenamefont
		{Stocker}}]{StokerReviewPlanktons}%
	\BibitemOpen
	\bibfield  {author} {\bibinfo {author} {\bibfnamefont {J.~S.}\ \bibnamefont
			{Guasto}}, \bibinfo {author} {\bibfnamefont {R.}~\bibnamefont {Rusconi}},\
		and\ \bibinfo {author} {\bibfnamefont {R.}~\bibnamefont {Stocker}},\
	}\bibfield  {title} {\bibinfo {title} {Fluid mechanics of planktonic
			microorganisms},\ }\href
	{https://doi.org/10.1146/annurev-fluid-120710-101156} {\bibfield  {journal}
		{\bibinfo  {journal} {Annu. Rev. Fluid Mech.}\ }\textbf {\bibinfo {volume}
			{44}},\ \bibinfo {pages} {373} (\bibinfo {year} {2012})}\BibitemShut
	{NoStop}%
	\bibitem [{\citenamefont {Pedley}\ and\ \citenamefont
		{Kessler}(1992)}]{PedleyReview}%
	\BibitemOpen
	\bibfield  {author} {\bibinfo {author} {\bibfnamefont {T.~J.}\ \bibnamefont
			{Pedley}}\ and\ \bibinfo {author} {\bibfnamefont {J.~O.}\ \bibnamefont
			{Kessler}},\ }\bibfield  {title} {\bibinfo {title} {Hydrodynamic phenomena in
			suspensions of swimming microorganisms},\ }\href
	{https://doi.org/10.1146/annurev.fl.24.010192.001525} {\bibfield  {journal}
		{\bibinfo  {journal} {Annu. Rev. Fluid Mech.}\ }\textbf {\bibinfo {volume}
			{24}},\ \bibinfo {pages} {313} (\bibinfo {year} {1992})}\BibitemShut
	{NoStop}%
	\bibitem [{\citenamefont {Rusconi}\ and\ \citenamefont
		{Stocker}(2015)}]{rusconi2015microbes}%
	\BibitemOpen
	\bibfield  {author} {\bibinfo {author} {\bibfnamefont {R.}~\bibnamefont
			{Rusconi}}\ and\ \bibinfo {author} {\bibfnamefont {R.}~\bibnamefont
			{Stocker}},\ }\bibfield  {title} {\bibinfo {title} {Microbes in flow},\
	}\href {https://doi.org/10.1016/j.mib.2015.03.003} {\bibfield  {journal}
		{\bibinfo  {journal} {Curr. Op. Microbiol.}\ }\textbf {\bibinfo {volume}
			{25}},\ \bibinfo {pages} {1} (\bibinfo {year} {2015})}\BibitemShut {NoStop}%
	\bibitem [{\citenamefont {Kessler}(1985)}]{Kessler1985}%
	\BibitemOpen
	\bibfield  {author} {\bibinfo {author} {\bibfnamefont {J.~O.}\ \bibnamefont
			{Kessler}},\ }\bibfield  {title} {\bibinfo {title} {Hydrodynamic focusing of
			motile algal cells},\ }\href {https://doi.org/10.1038/313218a0} {\bibfield
		{journal} {\bibinfo  {journal} {Nature}\ }\textbf {\bibinfo {volume} {313}},\
		\bibinfo {pages} {218} (\bibinfo {year} {1985})}\BibitemShut {NoStop}%
	\bibitem [{\citenamefont {Pedley}\ and\ \citenamefont
		{Kessler}(1990)}]{pedley_kessler_1990}%
	\BibitemOpen
	\bibfield  {author} {\bibinfo {author} {\bibfnamefont {T.~J.}\ \bibnamefont
			{Pedley}}\ and\ \bibinfo {author} {\bibfnamefont {J.~O.}\ \bibnamefont
			{Kessler}},\ }\bibfield  {title} {\bibinfo {title} {A new continuum model for
			suspensions of gyrotactic micro-organisms},\ }\href
	{https://doi.org/10.1017/S0022112090001914} {\bibfield  {journal} {\bibinfo
			{journal} {J. Fluid Mech.}\ }\textbf {\bibinfo {volume} {212}},\ \bibinfo
		{pages} {155–182} (\bibinfo {year} {1990})}\BibitemShut {NoStop}%
	\bibitem [{\citenamefont {Lillo}\ \emph {et~al.}(2014)\citenamefont {Lillo},
		\citenamefont {Cencini}, \citenamefont {Durham}, \citenamefont {Barry},
		\citenamefont {Stocker}, \citenamefont {Climent},\ and\ \citenamefont
		{Boffetta}}]{GyroTurbulence1}%
	\BibitemOpen
	\bibfield  {author} {\bibinfo {author} {\bibfnamefont {F.~D.}\ \bibnamefont
			{Lillo}}, \bibinfo {author} {\bibfnamefont {M.}~\bibnamefont {Cencini}},
		\bibinfo {author} {\bibfnamefont {W.~M.}\ \bibnamefont {Durham}}, \bibinfo
		{author} {\bibfnamefont {M.}~\bibnamefont {Barry}}, \bibinfo {author}
		{\bibfnamefont {R.}~\bibnamefont {Stocker}}, \bibinfo {author} {\bibfnamefont
			{E.}~\bibnamefont {Climent}},\ and\ \bibinfo {author} {\bibfnamefont
			{G.}~\bibnamefont {Boffetta}},\ }\bibfield  {title} {\bibinfo {title}
		{Turbulent fluid acceleration generates clusters of gyrotactic
			microorganisms},\ }\href {https://dx.doi.org/10.1103/PhysRevLett.112.044502}
	{\bibfield  {journal} {\bibinfo  {journal} {Phys. Rev. Lett.}\ }\textbf
		{\bibinfo {volume} {112}},\ \bibinfo {pages} {044502} (\bibinfo {year}
		{2014})}\BibitemShut {NoStop}%
	\bibitem [{\citenamefont {Santamaria}\ \emph {et~al.}(2014)\citenamefont
		{Santamaria}, \citenamefont {Lillo}, \citenamefont {Cencini},\ and\
		\citenamefont {Boffetta}}]{GyroTurbulence2}%
	\BibitemOpen
	\bibfield  {author} {\bibinfo {author} {\bibfnamefont {F.}~\bibnamefont
			{Santamaria}}, \bibinfo {author} {\bibfnamefont {F.~D.}\ \bibnamefont
			{Lillo}}, \bibinfo {author} {\bibfnamefont {M.}~\bibnamefont {Cencini}},\
		and\ \bibinfo {author} {\bibfnamefont {G.}~\bibnamefont {Boffetta}},\
	}\bibfield  {title} {\bibinfo {title} {Gyrotactic trapping in laminar and
			turbulent {Kolmogorov} flow},\ }\href {https://dx.doi.org/10.1063/1.4900956}
	{\bibfield  {journal} {\bibinfo  {journal} {Phys. Fluids}\ }\textbf {\bibinfo
			{volume} {26}},\ \bibinfo {pages} {111901} (\bibinfo {year}
		{2014})}\BibitemShut {NoStop}%
	\bibitem [{\citenamefont {Gustavsson}\ \emph {et~al.}(2016)\citenamefont
		{Gustavsson}, \citenamefont {Berglund}, \citenamefont {Jonsson},\ and\
		\citenamefont {Mehlig}}]{GyroTurbulence3}%
	\BibitemOpen
	\bibfield  {author} {\bibinfo {author} {\bibfnamefont {K.}~\bibnamefont
			{Gustavsson}}, \bibinfo {author} {\bibfnamefont {F.}~\bibnamefont
			{Berglund}}, \bibinfo {author} {\bibfnamefont {P.~R.}\ \bibnamefont
			{Jonsson}},\ and\ \bibinfo {author} {\bibfnamefont {B.}~\bibnamefont
			{Mehlig}},\ }\bibfield  {title} {\bibinfo {title} {Preferential sampling and
			small-scale clustering of gyrotactic microswimmers in turbulence},\ }\href
	{https://dx.doi.org/10.1103/PhysRevLett.116.108104} {\bibfield  {journal}
		{\bibinfo  {journal} {Phys. Rev. Lett.}\ }\textbf {\bibinfo {volume} {116}},\
		\bibinfo {pages} {108104} (\bibinfo {year} {2016})}\BibitemShut {NoStop}%
	\bibitem [{\citenamefont {Kaya}\ and\ \citenamefont
		{Koser}(2009)}]{EcoliJeffrey}%
	\BibitemOpen
	\bibfield  {author} {\bibinfo {author} {\bibfnamefont {T.}~\bibnamefont
			{Kaya}}\ and\ \bibinfo {author} {\bibfnamefont {H.}~\bibnamefont {Koser}},\
	}\bibfield  {title} {\bibinfo {title} {Characterization of hydrodynamic
			surface interactions of \emph{Escherichia coli} cell bodies in shear flow},\
	}\href {https://doi.org/10.1103/PhysRevLett.103.138103} {\bibfield  {journal}
		{\bibinfo  {journal} {Phys. Rev. Lett.}\ }\textbf {\bibinfo {volume} {103}},\
		\bibinfo {pages} {138103} (\bibinfo {year} {2009})}\BibitemShut {NoStop}%
	\bibitem [{\citenamefont {Kaya}\ and\ \citenamefont
		{Koser}(2012)}]{EcoliRheotaxis}%
	\BibitemOpen
	\bibfield  {author} {\bibinfo {author} {\bibfnamefont {T.}~\bibnamefont
			{Kaya}}\ and\ \bibinfo {author} {\bibfnamefont {H.}~\bibnamefont {Koser}},\
	}\bibfield  {title} {\bibinfo {title} {Direct upstream motility in
			\emph{Escherichia coli}},\ }\href
	{https://doi.org/https://doi.org/10.1016/j.bpj.2012.03.001} {\bibfield
		{journal} {\bibinfo  {journal} {Biophys. J.}\ }\textbf {\bibinfo {volume}
			{102}},\ \bibinfo {pages} {1514} (\bibinfo {year} {2012})}\BibitemShut
	{NoStop}%
	\bibitem [{\citenamefont {Marcos}\ \emph {et~al.}(2012)\citenamefont {Marcos},
		\citenamefont {Fu}, \citenamefont {Powers},\ and\ \citenamefont
		{Stocker}}]{BacterialRheotaxis}%
	\BibitemOpen
	\bibfield  {author} {\bibinfo {author} {\bibnamefont {Marcos}}, \bibinfo
		{author} {\bibfnamefont {H.~C.}\ \bibnamefont {Fu}}, \bibinfo {author}
		{\bibfnamefont {T.~R.}\ \bibnamefont {Powers}},\ and\ \bibinfo {author}
		{\bibfnamefont {R.}~\bibnamefont {Stocker}},\ }\bibfield  {title} {\bibinfo
		{title} {Bacterial rheotaxis},\ }\href
	{https://doi.org/10.1073/pnas.1120955109} {\bibfield  {journal} {\bibinfo
			{journal} {Proc. Natl. Acad. Sci. U.S.A.}\ }\textbf {\bibinfo {volume}
			{109}},\ \bibinfo {pages} {4780} (\bibinfo {year} {2012})}\BibitemShut
	{NoStop}%
	\bibitem [{\citenamefont {Figueroa-Morales}\ \emph {et~al.}(2015)\citenamefont
		{Figueroa-Morales}, \citenamefont {Leonardo~Miño}, \citenamefont {Rivera},
		\citenamefont {Caballero}, \citenamefont {Clément}, \citenamefont
		{Altshuler},\ and\ \citenamefont {Lindner}}]{EcoliAnkeLindner}%
	\BibitemOpen
	\bibfield  {author} {\bibinfo {author} {\bibfnamefont {N.}~\bibnamefont
			{Figueroa-Morales}}, \bibinfo {author} {\bibfnamefont {G.}~\bibnamefont
			{Leonardo~Miño}}, \bibinfo {author} {\bibfnamefont {A.}~\bibnamefont
			{Rivera}}, \bibinfo {author} {\bibfnamefont {R.}~\bibnamefont {Caballero}},
		\bibinfo {author} {\bibfnamefont {E.}~\bibnamefont {Clément}}, \bibinfo
		{author} {\bibfnamefont {E.}~\bibnamefont {Altshuler}},\ and\ \bibinfo
		{author} {\bibfnamefont {A.}~\bibnamefont {Lindner}},\ }\bibfield  {title}
	{\bibinfo {title} {Living on the edge: transfer and traffic of \emph{E. Coli}
			in a confined flow},\ }\href {https://doi.org/10.1039/C5SM00939A} {\bibfield
		{journal} {\bibinfo  {journal} {Soft Matter}\ }\textbf {\bibinfo {volume}
			{11}},\ \bibinfo {pages} {6284} (\bibinfo {year} {2015})}\BibitemShut
	{NoStop}%
	\bibitem [{\citenamefont {Dehkharghani}\ \emph {et~al.}(2019)\citenamefont
		{Dehkharghani}, \citenamefont {Waisbord}, \citenamefont {Dunkel},\ and\
		\citenamefont {Guasto}}]{BacterialCrystalDispersion}%
	\BibitemOpen
	\bibfield  {author} {\bibinfo {author} {\bibfnamefont {A.}~\bibnamefont
			{Dehkharghani}}, \bibinfo {author} {\bibfnamefont {N.}~\bibnamefont
			{Waisbord}}, \bibinfo {author} {\bibfnamefont {J.}~\bibnamefont {Dunkel}},\
		and\ \bibinfo {author} {\bibfnamefont {J.~S.}\ \bibnamefont {Guasto}},\
	}\bibfield  {title} {\bibinfo {title} {Bacterial scattering in microfluidic
			crystal flows reveals giant active {Taylor{\textendash}Aris} dispersion},\
	}\href {https://doi.org/10.1073/pnas.1819613116} {\bibfield  {journal}
		{\bibinfo  {journal} {Proc. Natl. Acad. Sci. U.S.A.}\ }\textbf {\bibinfo
			{volume} {116}},\ \bibinfo {pages} {11119} (\bibinfo {year}
		{2019})}\BibitemShut {NoStop}%
	\bibitem [{\citenamefont {Roberts}(1970)}]{sperm1}%
	\BibitemOpen
	\bibfield  {author} {\bibinfo {author} {\bibfnamefont {A.~M.}\ \bibnamefont
			{Roberts}},\ }\bibfield  {title} {\bibinfo {title} {Motion of spermatozoa in
			fluid streams},\ }\href {https://doi.org/10.1038/228375a0} {\bibfield
		{journal} {\bibinfo  {journal} {Nature}\ }\textbf {\bibinfo {volume} {228}},\
		\bibinfo {pages} {375} (\bibinfo {year} {1970})}\BibitemShut {NoStop}%
	\bibitem [{\citenamefont {Kantsler}\ \emph {et~al.}(2014)\citenamefont
		{Kantsler}, \citenamefont {Dunkel}, \citenamefont {Blayney},\ and\
		\citenamefont {Goldstein}}]{SpermRayRheot}%
	\BibitemOpen
	\bibfield  {author} {\bibinfo {author} {\bibfnamefont {V.}~\bibnamefont
			{Kantsler}}, \bibinfo {author} {\bibfnamefont {J.}~\bibnamefont {Dunkel}},
		\bibinfo {author} {\bibfnamefont {M.}~\bibnamefont {Blayney}},\ and\ \bibinfo
		{author} {\bibfnamefont {R.~E.}\ \bibnamefont {Goldstein}},\ }\bibfield
	{title} {\bibinfo {title} {{Rheotaxis facilitates upstream navigation of
				mammalian sperm cells}},\ }\href {https://doi.org/10.7554/eLife.02403.001}
	{\bibfield  {journal} {\bibinfo  {journal} {{eLife}}\ }\textbf {\bibinfo
			{volume} {{3}}} (\bibinfo {year} {{2014}})}\BibitemShut {NoStop}%
	\bibitem [{\citenamefont {Kumar}\ and\ \citenamefont
		{Ardekani}(2019)}]{SpermShear}%
	\BibitemOpen
	\bibfield  {author} {\bibinfo {author} {\bibfnamefont {M.}~\bibnamefont
			{Kumar}}\ and\ \bibinfo {author} {\bibfnamefont {A.~M.}\ \bibnamefont
			{Ardekani}},\ }\bibfield  {title} {\bibinfo {title} {Effect of external shear
			flow on sperm motility},\ }\href {https://doi.org/10.1039/C9SM00717B}
	{\bibfield  {journal} {\bibinfo  {journal} {Soft Matter}\ }\textbf {\bibinfo
			{volume} {15}},\ \bibinfo {pages} {6269} (\bibinfo {year}
		{2019})}\BibitemShut {NoStop}%
	\bibitem [{\citenamefont {Fauci}\ and\ \citenamefont
		{Dillon}(2006{\natexlab{b}})}]{SpermReview}%
	\BibitemOpen
	\bibfield  {author} {\bibinfo {author} {\bibfnamefont {L.~J.}\ \bibnamefont
			{Fauci}}\ and\ \bibinfo {author} {\bibfnamefont {R.}~\bibnamefont {Dillon}},\
	}\bibfield  {title} {\bibinfo {title} {Biofluidmechanics of reproduction},\
	}\href {https://doi.org/10.1146/annurev.fluid.37.061903.175725} {\bibfield
		{journal} {\bibinfo  {journal} {Annu. Rev. Fluid Mech.}\ }\textbf {\bibinfo
			{volume} {38}},\ \bibinfo {pages} {371} (\bibinfo {year}
		{2006}{\natexlab{b}})}\BibitemShut {NoStop}%
	\bibitem [{\citenamefont {Miki}\ and\ \citenamefont
		{Clapham}(2013)}]{miki2013rheotaxis}%
	\BibitemOpen
	\bibfield  {author} {\bibinfo {author} {\bibfnamefont {K.}~\bibnamefont
			{Miki}}\ and\ \bibinfo {author} {\bibfnamefont {D.~E.}\ \bibnamefont
			{Clapham}},\ }\bibfield  {title} {\bibinfo {title} {Rheotaxis guides
			mammalian sperm},\ }\href {https://doi.org/10.1016/j.cub.2013.02.007}
	{\bibfield  {journal} {\bibinfo  {journal} {Curr. Biol.}\ }\textbf {\bibinfo
			{volume} {23}},\ \bibinfo {pages} {443} (\bibinfo {year} {2013})}\BibitemShut
	{NoStop}%
	\bibitem [{\citenamefont {Kim}\ and\ \citenamefont
		{Julius}(2012)}]{MicroNanoBook}%
	\BibitemOpen
	\bibfield  {author} {\bibinfo {author} {\bibfnamefont {M.}~\bibnamefont
			{Kim}}\ and\ \bibinfo {author} {\bibfnamefont {A.~A.}\ \bibnamefont
			{Julius}},\ }\href@noop {} {\emph {\bibinfo {title} {Micro and Nano
				Technologies}}}\ (\bibinfo  {publisher} {Elsevier},\ \bibinfo {year}
	{2012})\BibitemShut {NoStop}%
	\bibitem [{\citenamefont {Palacci}\ \emph {et~al.}(2015)\citenamefont
		{Palacci}, \citenamefont {Sacanna}, \citenamefont {Abramian}, \citenamefont
		{Barral}, \citenamefont {Hanson}, \citenamefont {Grosberg}, \citenamefont
		{Pine},\ and\ \citenamefont {Chaikin}}]{ArtificialRheotaxis}%
	\BibitemOpen
	\bibfield  {author} {\bibinfo {author} {\bibfnamefont {J.}~\bibnamefont
			{Palacci}}, \bibinfo {author} {\bibfnamefont {S.}~\bibnamefont {Sacanna}},
		\bibinfo {author} {\bibfnamefont {A.}~\bibnamefont {Abramian}}, \bibinfo
		{author} {\bibfnamefont {J.}~\bibnamefont {Barral}}, \bibinfo {author}
		{\bibfnamefont {K.}~\bibnamefont {Hanson}}, \bibinfo {author} {\bibfnamefont
			{A.~Y.}\ \bibnamefont {Grosberg}}, \bibinfo {author} {\bibfnamefont {D.~J.}\
			\bibnamefont {Pine}},\ and\ \bibinfo {author} {\bibfnamefont {P.~M.}\
			\bibnamefont {Chaikin}},\ }\bibfield  {title} {\bibinfo {title} {Artificial
			rheotaxis},\ }\href {https://doi.org/10.1126/sciadv.1400214} {\bibfield
		{journal} {\bibinfo  {journal} {Sci. Adv.}\ }\textbf {\bibinfo {volume}
			{1}},\ \bibinfo {pages} {e1400214} (\bibinfo {year} {2015})}\BibitemShut
	{NoStop}%
	\bibitem [{\citenamefont {Baker}\ \emph {et~al.}(2019)\citenamefont {Baker},
		\citenamefont {Kauffman}, \citenamefont {Laskar}, \citenamefont {Shklyaev},
		\citenamefont {Potomkin}, \citenamefont {Dominguez-Rubio}, \citenamefont
		{Shum}, \citenamefont {Cruz-rivera}, \citenamefont {Aranson}, \citenamefont
		{Balazs},\ and\ \citenamefont {Sen}}]{ArtificialRodRheotaxis}%
	\BibitemOpen
	\bibfield  {author} {\bibinfo {author} {\bibfnamefont {R.}~\bibnamefont
			{Baker}}, \bibinfo {author} {\bibfnamefont {J.~E.}\ \bibnamefont {Kauffman}},
		\bibinfo {author} {\bibfnamefont {A.}~\bibnamefont {Laskar}}, \bibinfo
		{author} {\bibfnamefont {O.~E.}\ \bibnamefont {Shklyaev}}, \bibinfo {author}
		{\bibfnamefont {M.}~\bibnamefont {Potomkin}}, \bibinfo {author}
		{\bibfnamefont {L.}~\bibnamefont {Dominguez-Rubio}}, \bibinfo {author}
		{\bibfnamefont {H.}~\bibnamefont {Shum}}, \bibinfo {author} {\bibfnamefont
			{Y.}~\bibnamefont {Cruz-rivera}}, \bibinfo {author} {\bibfnamefont {I.~S.}\
			\bibnamefont {Aranson}}, \bibinfo {author} {\bibfnamefont {A.~C.}\
			\bibnamefont {Balazs}},\ and\ \bibinfo {author} {\bibfnamefont
			{A.}~\bibnamefont {Sen}},\ }\bibfield  {title} {\bibinfo {title} {Fight the
			flow: the role of shear in artificial rheotaxis for individual and collective
			motion},\ }\href {https://doi.org/10.1039/C8NR10257K} {\bibfield  {journal}
		{\bibinfo  {journal} {Nanoscale}\ }\textbf {\bibinfo {volume} {11}},\
		\bibinfo {pages} {10944} (\bibinfo {year} {2019})}\BibitemShut {NoStop}%
	\bibitem [{\citenamefont {Ren}\ \emph {et~al.}(2017)\citenamefont {Ren},
		\citenamefont {Zhou}, \citenamefont {Mao}, \citenamefont {Xu}, \citenamefont
		{Huang},\ and\ \citenamefont {Mallouk}}]{RheotaxisBimetalicMicro}%
	\BibitemOpen
	\bibfield  {author} {\bibinfo {author} {\bibfnamefont {L.}~\bibnamefont
			{Ren}}, \bibinfo {author} {\bibfnamefont {D.}~\bibnamefont {Zhou}}, \bibinfo
		{author} {\bibfnamefont {Z.}~\bibnamefont {Mao}}, \bibinfo {author}
		{\bibfnamefont {P.}~\bibnamefont {Xu}}, \bibinfo {author} {\bibfnamefont
			{T.~J.}\ \bibnamefont {Huang}},\ and\ \bibinfo {author} {\bibfnamefont
			{T.~E.}\ \bibnamefont {Mallouk}},\ }\bibfield  {title} {\bibinfo {title}
		{Rheotaxis of bimetallic micromotors driven by chemical--acoustic hybrid
			power},\ }\href {https://doi.org/10.1021/acsnano.7b06107} {\bibfield
		{journal} {\bibinfo  {journal} {Acs Nano}\ }\textbf {\bibinfo {volume}
			{11}},\ \bibinfo {pages} {10591} (\bibinfo {year} {2017})}\BibitemShut
	{NoStop}%
	\bibitem [{\citenamefont {Potomkin}\ \emph {et~al.}(2017)\citenamefont
		{Potomkin}, \citenamefont {Kaiser}, \citenamefont {Berlyand},\ and\
		\citenamefont {Aranson}}]{Potomkin_2017}%
	\BibitemOpen
	\bibfield  {author} {\bibinfo {author} {\bibfnamefont {M.}~\bibnamefont
			{Potomkin}}, \bibinfo {author} {\bibfnamefont {A.}~\bibnamefont {Kaiser}},
		\bibinfo {author} {\bibfnamefont {L.}~\bibnamefont {Berlyand}},\ and\
		\bibinfo {author} {\bibfnamefont {I.}~\bibnamefont {Aranson}},\ }\bibfield
	{title} {\bibinfo {title} {Focusing of active particles in a converging
			flow},\ }\href {https://doi.org/10.1088/1367-2630/aa94fd} {\bibfield
		{journal} {\bibinfo  {journal} {New J. Phys.}\ }\textbf {\bibinfo {volume}
			{19}},\ \bibinfo {pages} {115005} (\bibinfo {year} {2017})}\BibitemShut
	{NoStop}%
	\bibitem [{\citenamefont {Uspal}\ \emph {et~al.}(2015)\citenamefont {Uspal},
		\citenamefont {Popescu}, \citenamefont {Dietrich},\ and\ \citenamefont
		{Tasinkevych}}]{RheotaxisArtificialSphere}%
	\BibitemOpen
	\bibfield  {author} {\bibinfo {author} {\bibfnamefont {W.~E.}\ \bibnamefont
			{Uspal}}, \bibinfo {author} {\bibfnamefont {M.~N.}\ \bibnamefont {Popescu}},
		\bibinfo {author} {\bibfnamefont {S.}~\bibnamefont {Dietrich}},\ and\
		\bibinfo {author} {\bibfnamefont {M.}~\bibnamefont {Tasinkevych}},\
	}\bibfield  {title} {\bibinfo {title} {Rheotaxis of spherical active
			particles near a planar wall},\ }\href {https://doi.org/10.1039/C5SM01088H}
	{\bibfield  {journal} {\bibinfo  {journal} {Soft Matter}\ }\textbf {\bibinfo
			{volume} {11}},\ \bibinfo {pages} {6613} (\bibinfo {year}
		{2015})}\BibitemShut {NoStop}%
	\bibitem [{\citenamefont {Hatwalne}\ \emph {et~al.}(2004)\citenamefont
		{Hatwalne}, \citenamefont {Ramaswamy}, \citenamefont {Rao},\ and\
		\citenamefont {Simha}}]{RamaswamyActiveParticles}%
	\BibitemOpen
	\bibfield  {author} {\bibinfo {author} {\bibfnamefont {Y.}~\bibnamefont
			{Hatwalne}}, \bibinfo {author} {\bibfnamefont {S.}~\bibnamefont {Ramaswamy}},
		\bibinfo {author} {\bibfnamefont {M.}~\bibnamefont {Rao}},\ and\ \bibinfo
		{author} {\bibfnamefont {R.~A.}\ \bibnamefont {Simha}},\ }\bibfield  {title}
	{\bibinfo {title} {Rheology of active-particle suspensions},\ }\href
	{https://doi.org/10.1103/PhysRevLett.92.118101} {\bibfield  {journal}
		{\bibinfo  {journal} {Phys. Rev. Lett.}\ }\textbf {\bibinfo {volume} {92}},\
		\bibinfo {pages} {118101} (\bibinfo {year} {2004})}\BibitemShut {NoStop}%
	\bibitem [{\citenamefont {Torney}\ and\ \citenamefont
		{Neufeld}(2007)}]{TorneyNeufeld2007}%
	\BibitemOpen
	\bibfield  {author} {\bibinfo {author} {\bibfnamefont {C.}~\bibnamefont
			{Torney}}\ and\ \bibinfo {author} {\bibfnamefont {Z.}~\bibnamefont
			{Neufeld}},\ }\bibfield  {title} {\bibinfo {title} {Transport and aggregation
			of self-propelled particles in fluid flows},\ }\href
	{https://doi.org/10.1103/PhysRevLett.99.078101} {\bibfield  {journal}
		{\bibinfo  {journal} {Phys. Rev. Lett.}\ }\textbf {\bibinfo {volume} {99}},\
		\bibinfo {pages} {078101} (\bibinfo {year} {2007})}\BibitemShut {NoStop}%
	\bibitem [{\citenamefont {Hagen}\ \emph {et~al.}(2011)\citenamefont {Hagen},
		\citenamefont {Teeffelen},\ and\ \citenamefont {Löwen}}]{ten_Hagen_2011}%
	\BibitemOpen
	\bibfield  {author} {\bibinfo {author} {\bibfnamefont {B.~T.}\ \bibnamefont
			{Hagen}}, \bibinfo {author} {\bibfnamefont {S.~V.}\ \bibnamefont
			{Teeffelen}},\ and\ \bibinfo {author} {\bibfnamefont {H.}~\bibnamefont
			{Löwen}},\ }\bibfield  {title} {\bibinfo {title} {Brownian motion of a
			self-propelled particle},\ }\href
	{https://doi.org/10.1088/0953-8984/23/19/194119} {\bibfield  {journal}
		{\bibinfo  {journal} {J. Phys. Condens. Matter}\ }\textbf {\bibinfo {volume}
			{23}},\ \bibinfo {pages} {194119} (\bibinfo {year} {2011})}\BibitemShut
	{NoStop}%
	\bibitem [{\citenamefont {Stark}(2016)}]{Stark2016}%
	\BibitemOpen
	\bibfield  {author} {\bibinfo {author} {\bibfnamefont {H.}~\bibnamefont
			{Stark}},\ }\bibfield  {title} {\bibinfo {title} {Swimming in external
			fields},\ }\href {https://doi.org/10.1140/epjst/e2016-60060-2} {\bibfield
		{journal} {\bibinfo  {journal} {Eur. Phys. J. Spec. Top.}\ }\textbf {\bibinfo
			{volume} {225}},\ \bibinfo {pages} {2369} (\bibinfo {year}
		{2016})}\BibitemShut {NoStop}%
	\bibitem [{\citenamefont {Kim}\ and\ \citenamefont
		{Karrila}(2013)}]{KimKarrila}%
	\BibitemOpen
	\bibfield  {author} {\bibinfo {author} {\bibfnamefont {S.}~\bibnamefont
			{Kim}}\ and\ \bibinfo {author} {\bibfnamefont {S.~J.}\ \bibnamefont
			{Karrila}},\ }\href@noop {} {\emph {\bibinfo {title} {Microhydrodynamics:
				Principles and Selected Applications}}}\ (\bibinfo  {publisher} {Courier
		Corporation, North Chelmsford, MA},\ \bibinfo {year} {2013})\BibitemShut
	{NoStop}%
	\bibitem [{\citenamefont {Sandoval}\ \emph {et~al.}(2014)\citenamefont
		{Sandoval}, \citenamefont {Marath}, \citenamefont {Subramanian},\ and\
		\citenamefont {Lauga}}]{linearflow}%
	\BibitemOpen
	\bibfield  {author} {\bibinfo {author} {\bibfnamefont {M.}~\bibnamefont
			{Sandoval}}, \bibinfo {author} {\bibfnamefont {N.~K.}\ \bibnamefont
			{Marath}}, \bibinfo {author} {\bibfnamefont {G.}~\bibnamefont
			{Subramanian}},\ and\ \bibinfo {author} {\bibfnamefont {E.}~\bibnamefont
			{Lauga}},\ }\bibfield  {title} {\bibinfo {title} {Stochastic dynamics of
			active swimmers in linear flows},\ }\href
	{https://doi.org/10.1017/jfm.2013.651} {\bibfield  {journal} {\bibinfo
			{journal} {J. Fluid Mech.}\ }\textbf {\bibinfo {volume} {742}},\ \bibinfo
		{pages} {50–70} (\bibinfo {year} {2014})}\BibitemShut {NoStop}%
	\bibitem [{\citenamefont {Z\"ottl}\ and\ \citenamefont
		{Stark}(2012)}]{poisseuille}%
	\BibitemOpen
	\bibfield  {author} {\bibinfo {author} {\bibfnamefont {A.}~\bibnamefont
			{Z\"ottl}}\ and\ \bibinfo {author} {\bibfnamefont {H.}~\bibnamefont
			{Stark}},\ }\bibfield  {title} {\bibinfo {title} {Nonlinear dynamics of a
			microswimmer in {Poiseuille} flow},\ }\href
	{https://doi.org/10.1103/PhysRevLett.108.218104} {\bibfield  {journal}
		{\bibinfo  {journal} {Phys. Rev. Lett.}\ }\textbf {\bibinfo {volume} {108}},\
		\bibinfo {pages} {218104} (\bibinfo {year} {2012})}\BibitemShut {NoStop}%
	\bibitem [{\citenamefont {Jeffery}\ and\ \citenamefont
		{Filon}(1922)}]{Jeffery}%
	\BibitemOpen
	\bibfield  {author} {\bibinfo {author} {\bibfnamefont {G.~B.}\ \bibnamefont
			{Jeffery}}\ and\ \bibinfo {author} {\bibfnamefont {L.~N.~G.}\ \bibnamefont
			{Filon}},\ }\bibfield  {title} {\bibinfo {title} {The motion of ellipsoidal
			particles immersed in a viscous fluid},\ }\href
	{https://doi.org/10.1098/rspa.1922.0078} {\bibfield  {journal} {\bibinfo
			{journal} {Proc. R. Soc. Lond. A}\ }\textbf {\bibinfo {volume} {102}},\
		\bibinfo {pages} {161} (\bibinfo {year} {1922})}\BibitemShut {NoStop}%
	\bibitem [{\citenamefont {Binder}(1939)}]{cylindrical_particles}%
	\BibitemOpen
	\bibfield  {author} {\bibinfo {author} {\bibfnamefont {R.~C.}\ \bibnamefont
			{Binder}},\ }\bibfield  {title} {\bibinfo {title} {The motion of cylindrical
			particles in viscous flow},\ }\href {https://doi.org/10.1063/1.1707254}
	{\bibfield  {journal} {\bibinfo  {journal} {J. Appl. Phys.}\ }\textbf
		{\bibinfo {volume} {10}},\ \bibinfo {pages} {711} (\bibinfo {year}
		{1939})}\BibitemShut {NoStop}%
	\bibitem [{\citenamefont {Simha}(1940)}]{noise_elongated}%
	\BibitemOpen
	\bibfield  {author} {\bibinfo {author} {\bibfnamefont {R.}~\bibnamefont
			{Simha}},\ }\bibfield  {title} {\bibinfo {title} {The influence of brownian
			movement on the viscosity of solutions.},\ }\href
	{https://doi.org/10.1021/j150397a004} {\bibfield  {journal} {\bibinfo
			{journal} {J. Phys. Chem.}\ }\textbf {\bibinfo {volume} {44}},\ \bibinfo
		{pages} {25} (\bibinfo {year} {1940})}\BibitemShut {NoStop}%
	\bibitem [{\citenamefont {Leal}\ and\ \citenamefont
		{Hinch}(1971)}]{leal_hinch_1971_shear}%
	\BibitemOpen
	\bibfield  {author} {\bibinfo {author} {\bibfnamefont {L.~G.}\ \bibnamefont
			{Leal}}\ and\ \bibinfo {author} {\bibfnamefont {E.~J.}\ \bibnamefont
			{Hinch}},\ }\bibfield  {title} {\bibinfo {title} {The effect of weak
			{B}rownian rotations on particles in shear flow},\ }\href
	{https://doi.org/10.1017/S0022112071000788} {\bibfield  {journal} {\bibinfo
			{journal} {J. Fluid Mech.}\ }\textbf {\bibinfo {volume} {46}},\ \bibinfo
		{pages} {685–703} (\bibinfo {year} {1971})}\BibitemShut {NoStop}%
	\bibitem [{\citenamefont {Berman}\ \emph {et~al.}(2021)\citenamefont {Berman},
		\citenamefont {Buggeln}, \citenamefont {Brantley}, \citenamefont {Mitchell},\
		and\ \citenamefont {Solomon}}]{HyperbolicFlow}%
	\BibitemOpen
	\bibfield  {author} {\bibinfo {author} {\bibfnamefont {S.~A.}\ \bibnamefont
			{Berman}}, \bibinfo {author} {\bibfnamefont {J.}~\bibnamefont {Buggeln}},
		\bibinfo {author} {\bibfnamefont {D.~A.}\ \bibnamefont {Brantley}}, \bibinfo
		{author} {\bibfnamefont {K.~A.}\ \bibnamefont {Mitchell}},\ and\ \bibinfo
		{author} {\bibfnamefont {T.~H.}\ \bibnamefont {Solomon}},\ }\bibfield
	{title} {\bibinfo {title} {Transport barriers to self-propelled particles in
			fluid flows},\ }\href {https://doi.org/10.1103/PhysRevFluids.6.L012501}
	{\bibfield  {journal} {\bibinfo  {journal} {Phys. Rev. Fluids}\ }\textbf
		{\bibinfo {volume} {6}},\ \bibinfo {pages} {L012501} (\bibinfo {year}
		{2021})}\BibitemShut {NoStop}%
	\bibitem [{\citenamefont {Ezhilan}\ and\ \citenamefont
		{Saintillan}(2015)}]{PressureDriven}%
	\BibitemOpen
	\bibfield  {author} {\bibinfo {author} {\bibfnamefont {B.}~\bibnamefont
			{Ezhilan}}\ and\ \bibinfo {author} {\bibfnamefont {D.}~\bibnamefont
			{Saintillan}},\ }\bibfield  {title} {\bibinfo {title} {Transport of a dilute
			active suspension in pressure-driven channel flow},\ }\href
	{https://doi.org/10.1017/jfm.2015.372} {\bibfield  {journal} {\bibinfo
			{journal} {J. Fluid Mech.}\ }\textbf {\bibinfo {volume} {777}},\ \bibinfo
		{pages} {482–522} (\bibinfo {year} {2015})}\BibitemShut {NoStop}%
	\bibitem [{\citenamefont {Z{\"o}ttl}\ and\ \citenamefont
		{Stark}(2013)}]{zottl2013periodic}%
	\BibitemOpen
	\bibfield  {author} {\bibinfo {author} {\bibfnamefont {A.}~\bibnamefont
			{Z{\"o}ttl}}\ and\ \bibinfo {author} {\bibfnamefont {H.}~\bibnamefont
			{Stark}},\ }\bibfield  {title} {\bibinfo {title} {Periodic and quasiperiodic
			motion of an elongated microswimmer in {Poiseuille} flow},\ }\href
	{https://doi.org/10.1140/epje/i2013-13004-5} {\bibfield  {journal} {\bibinfo
			{journal} {Eur. Phys. J. E}\ }\textbf {\bibinfo {volume} {36}},\ \bibinfo
		{pages} {4} (\bibinfo {year} {2013})}\BibitemShut {NoStop}%
	\bibitem [{\citenamefont {Rusconi}\ \emph {et~al.}(2014)\citenamefont
		{Rusconi}, \citenamefont {Guasto},\ and\ \citenamefont
		{Stocker}}]{rusconi14}%
	\BibitemOpen
	\bibfield  {author} {\bibinfo {author} {\bibfnamefont {R.}~\bibnamefont
			{Rusconi}}, \bibinfo {author} {\bibfnamefont {J.~S.}\ \bibnamefont
			{Guasto}},\ and\ \bibinfo {author} {\bibfnamefont {R.}~\bibnamefont
			{Stocker}},\ }\bibfield  {title} {\bibinfo {title} {Bacterial transport
			suppressed by fluid shear},\ }\href {https://doi.org/10.1038/nphys2883}
	{\bibfield  {journal} {\bibinfo  {journal} {Nature Phys.}\ }\textbf {\bibinfo
			{volume} {10}},\ \bibinfo {pages} {212} (\bibinfo {year} {2014})}\BibitemShut
	{NoStop}%
	\bibitem [{\citenamefont {Khurana}\ \emph {et~al.}(2011)\citenamefont
		{Khurana}, \citenamefont {Blawzdziewicz},\ and\ \citenamefont
		{Ouellette}}]{khurana2011reduced}%
	\BibitemOpen
	\bibfield  {author} {\bibinfo {author} {\bibfnamefont {N.}~\bibnamefont
			{Khurana}}, \bibinfo {author} {\bibfnamefont {J.}~\bibnamefont
			{Blawzdziewicz}},\ and\ \bibinfo {author} {\bibfnamefont {N.~T.}\
			\bibnamefont {Ouellette}},\ }\bibfield  {title} {\bibinfo {title} {Reduced
			transport of swimming particles in chaotic flow due to hydrodynamic
			trapping},\ }\href {https://link.aps.org/doi/10.1103/PhysRevLett.106.198104}
	{\bibfield  {journal} {\bibinfo  {journal} {Phys. Rev. Lett.}\ }\textbf
		{\bibinfo {volume} {106}},\ \bibinfo {pages} {198104} (\bibinfo {year}
		{2011})}\BibitemShut {NoStop}%
	\bibitem [{\citenamefont {Khurana}\ and\ \citenamefont
		{Ouellette}(2012)}]{khurana2012interactions}%
	\BibitemOpen
	\bibfield  {author} {\bibinfo {author} {\bibfnamefont {N.}~\bibnamefont
			{Khurana}}\ and\ \bibinfo {author} {\bibfnamefont {N.~T.}\ \bibnamefont
			{Ouellette}},\ }\bibfield  {title} {\bibinfo {title} {Interactions between
			active particles and dynamical structures in chaotic flow},\ }\href
	{https://doi.org/10.1063/1.4754873} {\bibfield  {journal} {\bibinfo
			{journal} {Phys. Fluids}\ }\textbf {\bibinfo {volume} {24}},\ \bibinfo
		{pages} {091902} (\bibinfo {year} {2012})}\BibitemShut {NoStop}%
	\bibitem [{\citenamefont {Lazier}\ and\ \citenamefont
		{Mann}(1989)}]{TurbulenceVortex1}%
	\BibitemOpen
	\bibfield  {author} {\bibinfo {author} {\bibfnamefont {J.~R.~N.}\
			\bibnamefont {Lazier}}\ and\ \bibinfo {author} {\bibfnamefont {K.~H.}\
			\bibnamefont {Mann}},\ }\bibfield  {title} {\bibinfo {title} {Turbulence and
			the diffusive layers around small organisms},\ }\href
	{https://doi.org/https://doi.org/10.1016/0198-0149(89)90068-X} {\bibfield
		{journal} {\bibinfo  {journal} {Deep Sea Res. Part A Oceanogr. Res. Pap.}\
		}\textbf {\bibinfo {volume} {36}},\ \bibinfo {pages} {1721} (\bibinfo {year}
		{1989})}\BibitemShut {NoStop}%
	\bibitem [{\citenamefont {Borgnino}\ \emph {et~al.}(2019)\citenamefont
		{Borgnino}, \citenamefont {Gustavsson}, \citenamefont {De~Lillo},
		\citenamefont {Boffetta}, \citenamefont {Cencini},\ and\ \citenamefont
		{Mehlig}}]{TurbulenceVortex2}%
	\BibitemOpen
	\bibfield  {author} {\bibinfo {author} {\bibfnamefont {M.}~\bibnamefont
			{Borgnino}}, \bibinfo {author} {\bibfnamefont {K.}~\bibnamefont
			{Gustavsson}}, \bibinfo {author} {\bibfnamefont {F.}~\bibnamefont
			{De~Lillo}}, \bibinfo {author} {\bibfnamefont {G.}~\bibnamefont {Boffetta}},
		\bibinfo {author} {\bibfnamefont {M.}~\bibnamefont {Cencini}},\ and\ \bibinfo
		{author} {\bibfnamefont {B.}~\bibnamefont {Mehlig}},\ }\bibfield  {title}
	{\bibinfo {title} {Alignment of nonspherical active particles in chaotic
			flows},\ }\href {https://doi.org/10.1103/PhysRevLett.123.138003} {\bibfield
		{journal} {\bibinfo  {journal} {Phys. Rev. Lett.}\ }\textbf {\bibinfo
			{volume} {123}},\ \bibinfo {pages} {138003} (\bibinfo {year}
		{2019})}\BibitemShut {NoStop}%
	\bibitem [{\citenamefont {Marcos}\ and\ \citenamefont
		{Stocker}(2006)}]{cavity}%
	\BibitemOpen
	\bibfield  {author} {\bibinfo {author} {\bibnamefont {Marcos}}\ and\ \bibinfo
		{author} {\bibfnamefont {R.}~\bibnamefont {Stocker}},\ }\bibfield  {title}
	{\bibinfo {title} {Microorganisms in vortices: a microfluidic setup},\ }\href
	{https://doi.org/10.4319/lom.2006.4.392} {\bibfield  {journal} {\bibinfo
			{journal} {Limnol. Oceanogr. Meth.}\ }\textbf {\bibinfo {volume} {4}},\
		\bibinfo {pages} {392} (\bibinfo {year} {2006})}\BibitemShut {NoStop}%
	\bibitem [{\citenamefont {Li}\ \emph {et~al.}(2021)\citenamefont {Li},
		\citenamefont {Lin}, \citenamefont {Ho}, \citenamefont {Hsieh},\ and\
		\citenamefont {Sun}}]{Li2021}%
	\BibitemOpen
	\bibfield  {author} {\bibinfo {author} {\bibfnamefont {S.-W.}\ \bibnamefont
			{Li}}, \bibinfo {author} {\bibfnamefont {P.-H.}\ \bibnamefont {Lin}},
		\bibinfo {author} {\bibfnamefont {T.-Y.}\ \bibnamefont {Ho}}, \bibinfo
		{author} {\bibfnamefont {C.-h.}\ \bibnamefont {Hsieh}},\ and\ \bibinfo
		{author} {\bibfnamefont {C.-l.}\ \bibnamefont {Sun}},\ }\bibfield  {title}
	{\bibinfo {title} {Change in rheotactic behavior patterns of dinoflagellates
			in response to different microfluidic environments},\ }\href
	{https://doi.org/10.1038/s41598-021-90622-8} {\bibfield  {journal} {\bibinfo
			{journal} {Sci. Rep.}\ }\textbf {\bibinfo {volume} {11}},\ \bibinfo {pages}
		{11105} (\bibinfo {year} {2021})}\BibitemShut {NoStop}%
	\bibitem [{\citenamefont {Sokolov}\ and\ \citenamefont
		{Aranson}(2016)}]{depletion}%
	\BibitemOpen
	\bibfield  {author} {\bibinfo {author} {\bibfnamefont {A.}~\bibnamefont
			{Sokolov}}\ and\ \bibinfo {author} {\bibfnamefont {I.~S.}\ \bibnamefont
			{Aranson}},\ }\bibfield  {title} {\bibinfo {title} {Rapid expulsion of
			microswimmers by a vortical flow},\ }\href
	{http://dx.doi.org/10.1038/ncomms11114} {\bibfield  {journal} {\bibinfo
			{journal} {Nat. Commun.}\ }\textbf {\bibinfo {volume} {7}},\ \bibinfo {pages}
		{11114 EP } (\bibinfo {year} {2016})}\BibitemShut {NoStop}%
	\bibitem [{\citenamefont {Arguedas-leiva}\ and\ \citenamefont
		{Wilczek}(2020)}]{Arguedas_Leiva_2020}%
	\BibitemOpen
	\bibfield  {author} {\bibinfo {author} {\bibfnamefont {J.~A.}\ \bibnamefont
			{Arguedas-leiva}}\ and\ \bibinfo {author} {\bibfnamefont {M.}~\bibnamefont
			{Wilczek}},\ }\bibfield  {title} {\bibinfo {title} {Microswimmers in an
			axisymmetric vortex flow},\ }\href {https://doi.org/10.1088/1367-2630/ab776f}
	{\bibfield  {journal} {\bibinfo  {journal} {New J. Phys.}\ }\textbf {\bibinfo
			{volume} {22}},\ \bibinfo {pages} {053051} (\bibinfo {year}
		{2020})}\BibitemShut {NoStop}%
	\bibitem [{\citenamefont {Velho~Rodrigues}\ \emph {et~al.}(2021)\citenamefont
		{Velho~Rodrigues}, \citenamefont {Lisicki},\ and\ \citenamefont
		{Lauga}}]{BOSO}%
	\BibitemOpen
	\bibfield  {author} {\bibinfo {author} {\bibfnamefont {M.~F.}\ \bibnamefont
			{Velho~Rodrigues}}, \bibinfo {author} {\bibfnamefont {M.}~\bibnamefont
			{Lisicki}},\ and\ \bibinfo {author} {\bibfnamefont {E.}~\bibnamefont
			{Lauga}},\ }\bibfield  {title} {\bibinfo {title} {The bank of swimming
			organisms at the micron scale (boso-micro)},\ }\href
	{https://doi.org/10.1371/journal.pone.0252291} {\bibfield  {journal}
		{\bibinfo  {journal} {PLOS ONE}\ }\textbf {\bibinfo {volume} {16}},\ \bibinfo
		{pages} {1} (\bibinfo {year} {2021})}\BibitemShut {NoStop}%
	\bibitem [{\citenamefont {Abramowitz}\ and\ \citenamefont
		{Stegun}(1964)}]{abramowitz1964handbook}%
	\BibitemOpen
	\bibfield  {author} {\bibinfo {author} {\bibfnamefont {M.}~\bibnamefont
			{Abramowitz}}\ and\ \bibinfo {author} {\bibfnamefont {I.~A.}\ \bibnamefont
			{Stegun}},\ }\href@noop {} {\emph {\bibinfo {title} {Handbook of mathematical
				functions with formulas, graphs, and mathematical tables}}},\ Vol.~\bibinfo
	{volume} {55}\ (\bibinfo  {publisher} {US Government printing office},\
	\bibinfo {year} {1964})\BibitemShut {NoStop}%
	\bibitem [{\citenamefont {Gardiner}\ \emph {et~al.}(1985)\citenamefont
		{Gardiner} \emph {et~al.}}]{FP_book}%
	\BibitemOpen
	\bibfield  {author} {\bibinfo {author} {\bibfnamefont {C.~W.}\ \bibnamefont
			{Gardiner}} \emph {et~al.},\ }\href@noop {} {\emph {\bibinfo {title}
			{Handbook of stochastic methods}}},\ Vol.~\bibinfo {volume} {3}\ (\bibinfo
	{publisher} {Springer, Berlin},\ \bibinfo {year} {1985})\BibitemShut
	{NoStop}%
	\bibitem [{\citenamefont {Siegert}(1951)}]{mfpt_paper}%
	\BibitemOpen
	\bibfield  {author} {\bibinfo {author} {\bibfnamefont {A.~J.~F.}\
			\bibnamefont {Siegert}},\ }\bibfield  {title} {\bibinfo {title} {On the first
			passage time probability problem},\ }\href
	{https://doi.org/10.1103/PhysRev.81.617} {\bibfield  {journal} {\bibinfo
			{journal} {Phys. Rev.}\ }\textbf {\bibinfo {volume} {81}},\ \bibinfo {pages}
		{617} (\bibinfo {year} {1951})}\BibitemShut {NoStop}%
	\bibitem [{\citenamefont {Redner}(2001)}]{mfpt_book}%
	\BibitemOpen
	\bibfield  {author} {\bibinfo {author} {\bibfnamefont {S.}~\bibnamefont
			{Redner}},\ }\href {https://doi.org/10.1017/CBO9780511606014} {\emph
		{\bibinfo {title} {A Guide to First-Passage Processes}}}\ (\bibinfo
	{publisher} {Cambridge University Press},\ \bibinfo {year}
	{2001})\BibitemShut {NoStop}%
	\bibitem [{\citenamefont {Hecht}(2012)}]{freefem}%
	\BibitemOpen
	\bibfield  {author} {\bibinfo {author} {\bibfnamefont {F.}~\bibnamefont
			{Hecht}},\ }\bibfield  {title} {\bibinfo {title} {New development in
			{FreeFem++}},\ }\href {https://freefem.org/} {\bibfield  {journal} {\bibinfo
			{journal} {J. Numer. Math.}\ }\textbf {\bibinfo {volume} {20}},\ \bibinfo
		{pages} {251} (\bibinfo {year} {2012})}\BibitemShut {NoStop}%
\end{thebibliography}
\end{document}